\documentclass{amsart}
\usepackage{amssymb}
\usepackage{amsfonts}

\newtheorem{theorem}{Theorem}
\theoremstyle{plain}

\numberwithin{equation}{section}
\input{tcilatex}

\begin{document}
\title[Quantum Diffusion, Measurement and Filtering]{Quantum Diffusion,
Measurement and Filtering}
\author{V P Belavkin}
\address{Moscow Institute of Electonics and Mathematics\\
Moscow 109028 USSR}
\email{vpb@maths.nott.ac.uk}
\urladdr{http://www.maths.nott.ac.uk/personal/vpb/}
\thanks{}
\date{April 30, 1991}
\dedicatory{}
\thanks{This paper was originally published in two parts in \textit{%
Probability Theory and its Applications} \textbf{38} pp 742--757, (1993) and 
\textbf{39} pp 640--658 (1994). }

\begin{abstract}
A brief presentation of the basic concepts in quantum probability theory is
given in comparison to the classical one. The notion of quantum white noise,
its explicit representation in Fock space, and necessary results of
noncommutative stochastic analysis and integration are outlined.

Algebraic differential equations that unify the quantum non Markovian
diffusion with continuous non demolition observation are derived. A
stochastic equation of quantum diffusion filtering generalising the
classical Markov filtering equation to the quantum flows over arbitrary
*-algebra is obtained.

A Gaussian quantum diffusion with one dimensional continuous observation is
considered.The a posteriori quantum state difusion in this case is reduced
to a linear quantum stochastic filter equation of Kalman-Bucy type and to
the operator Riccati equation for quantum correlations. An example of
continuous nondemolition observation of the coordinate of a free quantum
particle is considered, describing a continuous collase to the stationary
solution of the linear quantum filtering problem found in the paper.
\end{abstract}

\maketitle
\tableofcontents

\section{Introduction}

\setcounter{equation}{0}

Beginning in the mid seventies, modern probability theory has, (along with
traditional subjects, such as dynamical systems with random perturbations),
been also concerned with fundamentally new stochastic objects --- quantum
dynamic systems with an inherently probabilistic nature. Mathematically, the
concept of quantum probability arises not because of the lack of information
for a complete description of the object, the instability of chaotic motion
or the inaccuracy of measurement but is due to the noncommutativity of the
algebra of random variables which are represented by the operators in the
Hilbert space. As quantum probability theory is an intrinsically stochastic
theory, it cannot be stated within the framework of the Kolmogorov axioms 
\cite{bib:diff1}, which assume the fundamentally deterministic description
of the classical systems under the given point states $\omega \in \Omega $.
It is based on different axioms\cite{bib:diff2}, \cite{bib:diff3} namely,
the Neumann axioms, whose greater generality can be demostrated even in the
case of a finite number of alternative elementary events $\omega =1,\ldots
,n $.

Let us illustrate for this simple case how the classical probability space $%
(\Omega ,\mathcal{F},\mathbf{P})$ can be represented as a special case of
the quantum space that is defined by the triple $(\mathrm{H},\mathcal{A},%
\mathbf{E})$. Here \textrm{H} is a (finite-dimensional) complex space of
column-vectors $h=[\eta ^i]$, $\eta ^i\in \mathbf{C}$ with scalar product $%
(g|h)=\sum \overline{\zeta }^ip_i\eta ^i\equiv g^{*}h$ for $g=[\zeta ^i]\in 
\mathrm{H}$ defined by the weights (probabilities) $p_i>0$, $\mathcal{A}$ is
an associative, but not necessarily commutative matrix algebra $X=[\xi _k^i]$
closed under the involution $X\mapsto X^{*}$ defined by the Hermitian
conjugation%
\begin{equation*}
(X^{*}g|h)=(g|Xh),[\xi _k^i]^{*}=[p_i^{-1}\bar \xi _i^kp_k],
\end{equation*}
and with matrix $I=[\delta _k^i]$ as the identity $I\in \mathcal{A}$, and $%
\mathbf{E}[X]=(e|Xe)$ is a positive normilized functional ($\mathbf{E}%
[X^{*}X]\geq 0$, $\mathbf{E}[I]=1$) of expectation of noncommuting variables 
$X$ defined by a fixed unit vector $e\in \mathrm{H}$, where $\Vert e\Vert
^2:=(e|e)=1$.

Classical random variables $x:\Omega \to \mathbf{C}$ can also be described
by the multiplication operators $X=\widehat{x}$ , $(\widehat{x}h)(\omega
)=x(\omega )\,h(\omega )$ in the complex Hilbert space $\mathrm{H}%
=L^2(\Omega ,\mathcal{F},\mathbf{P})$ of $\mathcal{F}$ --measurable $\mathbf{%
P}$-square-integrable functions $h:\Omega \rightarrow \mathbf{C}$ , 
\begin{equation*}
\left\| h\right\| ^2=\int |h(\omega )|^2\mathbf{P}(\mathrm{d}\omega
)=(h|h)<\infty .
\end{equation*}
Their expectations $\mathbf{E}[\widehat{x}]=\int x(\omega )\,\mathbf{P}(%
\mathrm{d}\,\omega )$ are defined as $(e|\widehat{x}e)$ by the unit function 
$e(\omega )\equiv 1$ which is normalized with respect to any probability
measure $\mathbf{P}$. Thus, however, only commutative operator algebras $%
\mathcal{A}$ are obtained, whose elements are given in the finite case of $%
\Omega =\{1,\ldots ,n\}$ by all the diagonal matrices $\widehat{x}=[\xi
(i)\delta _k^i]$ with the commutative product, corresponding to the
pointwise multiplication of the functions $x(\omega )=\xi (i)$, where $%
\omega =i$.

Conversely, any quantum probability space $(\mathrm{H},\mathcal{A},\mathbf{E}%
)$ can be reduced to the classical one $(\Omega ,\mathcal{F},\mathbf{P})$
only in the case of the commutativity of the algebra $\mathcal{A}$; in the
finite-dimensional case this is realised by simultaneous reduction of the
commutting matrices $X\in \mathcal{A}$ to diagonal form $[\xi (i)\delta
_k^i] $ . The probabilities $p_i$ of the elementary events $\omega =i$ in
the diagonal representation are defined by the restriction $p_i=\mathbf{E}%
[P_i]$ of the functional $\mathbf{E}[X]=\sum \xi (i)p_i$ on projective
matrices $P_j=[\xi _j(i)\,\delta _i^k]$, $\xi _j(i)=\delta _j^i$.

In this article a quantum analog of diffusion and the problem of its
continuous measurement and stochastic filtering, that gives the solution of
the Zeno paradox \cite{bib:diff4} (as a result of establishing an a
posteriori stationary state), are considered within the framework of the
noncommutative algebraic approach. A derivation of the stochastic equation
is given for a nonnormalized a posteriori quantum state, which is obtained
in \cite{bib:diff5} by renormalising the basic equation of nonlinear quantum
filtering \cite{bib:diff6}. The solution of the equation has been found for
the case of linear quantum diffusion of canonical commutation relations,
obtained previously for the quantum Gaussian case by means of linear Markov
filtering methods in \cite{bib:diff7}, \cite{bib:diff8}.

In presenting the second (basic) section, we deliberately avoided the
questions concerning the sufficient conditions for the dense definition of
the unbounded infinitesimal generators that guarantee the uniqueness of
solutions of quantum stochastic and operator equations; this is beyond the
scope of this article. We only point out that in the first and second
sections these questions are not relevant (see \cite{bib:diff9}) for the
Markovian case with complete pre-Hilbert domain \textrm{D} in the initial
Hilbert space \textrm{H}, corresponding to boundedness of the operators $L$
and $H$ in $\mathrm{D=H}$. Moreover, a solution exists for an unbounded
algebra $\mathcal{A}$ of canonical commutation relations, which is
considered in the third section, in the framework of a quantum calculus of
kernels for the operators $L,$ $H\in \mathcal{A}$ in the initial Fock scale $%
\left\{ \mathrm{F}_\xi |\xi >1\right\} $\cite{bib:diff10} if their inductive
limit $\cup $\textrm{F}$_\xi $ is chosen as \textrm{D}. Besides, the
explicitly solvable model of this section with linear unbounded generators $%
L $ and $H$, does not require the estimates obtained in these scales.

For completeness the notation and explicite methods of quantum stochastic
integration and the proof of their estimates in Fock scale \cite{bib:diff5}
are briefly presented in the Appendix . The comprehensive statement of the
author's general approach, outlined above, and the estimates for the
integrals can be found in \cite{bib:diff10}, \cite{bib:diff11}. The earlier
results on quantum stochastic calculus in the framework of Hudson and
Parthasarathy approach \cite{bib:diff12}, are reviewed in\cite{bib:diff13}.

The approach presented generalizes the results for purely quantum diffusion
in \cite{bib:diff14} to the case of an arbitrary initial algebra $\mathcal{A}
$. This enables a unified description of quantum and classical diffusion,
their observation and filtering as special algebraic cases. In the sections
3 and 4, a one-dimensional variant of an infinite-dimensional quantum
Gaussian filtering \cite{bib:diff14} is presented as well as an example of
observation of a coordinate of a free quantum Brownian particle; this was
analysed earlier in \cite{bib:diff16} by the method of solving the a
posteriori Shr\"odinger equation \cite{bib:diff15}.

\section{Quantum diffusion and nondemolition measurement}

\setcounter{equation}{0}

\typeout{Quantum diffusion and nondemolition measurement}

\noindent \textbf{1.1. Basic Notation.} Let $\mathrm{H}$ be a complex
Hilbert space and $\mathrm{D}\subseteq \mathrm{H}$ be a dense subspace
defined as an inductive limit (see appendix 1) of some scale $\{\mathrm{H}%
_\xi |\xi >1\}$ in the space $\mathrm{H}$. Let the initial algebra $\mathcal{%
A}$ of noncommutative random variables describing a `quantum object' at the
initial moment $t=0$ be represented by an involutive subalgebra $\mathcal{A}%
\subseteq \mathcal{B}(\mathrm{D})$ of linear operators $X:\mathrm{D}\to 
\mathrm{D}$, $X^{*}\in \mathcal{A}$, having (an inductively) continuous
conjugate $X^{*}:\mathrm{D}\to \mathrm{D}$, $(X^{*}\chi |\psi )=(\chi |X\psi
)$ with respect to the scalar product in $\mathrm{H}$, with an identity
operator $I\in \mathcal{A}$.

Let us denote by $\mathcal{H}=\mathrm{H}\otimes \mathcal{F}$ the tensor
product $\mathrm{H}$ and the Fock space $\mathcal{F}=\Gamma (\mathcal{K})$
over the Hilbert space $\mathcal{K}=L^2(\mathbf{R}_{+})$ of a `quantum
noise' $\widehat{w}_t(g)$, $g\in \mathcal{K}$, and let the pre-Hilbert space 
$\mathcal{D}$ be an inductive limit of the Hilbert scale $\mathcal{H}_\xi =%
\mathrm{H}_\xi \otimes \mathcal{F}_\xi $, $\xi >1$, where $\{\mathcal{F}_\xi
\}$ is the natural Fock scale (see appendix 2) over $\mathcal{K}$. We shall
consider the quantum noise as a set $\{\widehat{w}_t(g)|g\in L^2(\mathbf{R}%
_{+})\}$ of Brownian motions $t\mapsto \widehat{w}_t(g) $, represented in $%
\mathcal{F}$ by self-adjoint operators 
\begin{equation*}
\widehat{w}_t(g)=\int_0^t\big(g(r)\,\mathrm{d}\,\widehat{a}_r^{*}+\bar g(r)\,%
\mathrm{d}\,\widehat{a}_r\big)\equiv \widehat{a}_t^{*}(g)+\widehat{a}%
_t(g^{*}),
\end{equation*}
with a Gaussian state on the algebra generated by them, which is induced by
the vacuum function $\delta _\emptyset \in \mathcal{F}$. Here $\{\widehat{a}%
_r,\widehat{a}_r^{*}|r\in \mathbf{R}_{+}\}$ are canonical operators of
creation $\widehat{a}_r^{*}$ and annihilation $\widehat{a}_r$ in $\mathcal{F}
$ (see Appendix 3) called quantum stochastic integrators, and $g^{*}(t)=%
\overline{g}(t)$ . Note that each operator function $t\mapsto \widehat{w}%
_t(g)$ that has commutative values $[\widehat{w}_s(g),\widehat{w}_t(g)]=0$
is equivalent to a classical Brownian motion with intensity $|g(t)|^2$, with
respect to the vacuum vector $e=\delta _\emptyset $. This follows from the
formula 
\begin{equation*}
e^{i\widehat{w}(g)}=e^{i\widehat{a}^{*}(g)}e^{-\frac 12\left\| g\right\|
^2}e^{i\widehat{a}(g^{*})}
\end{equation*}
and $e^{\widehat{a}(f)}\delta _\emptyset =\delta _\emptyset $ for any $f\in 
\mathcal{K}$ , due to which the quantum characteristic function 
\begin{equation*}
\mathbf{E}[e^{i\hat w_t(g)}]=(\delta _\emptyset |e^{i\hat w_t(g)}\delta
_\emptyset )
\end{equation*}
coincides with the clasical Gaussian characteristic function 
\begin{equation*}
\int \exp \{i\int_0^tg(r)\mathrm{d}w_r\}\mathbf{P}(\mathrm{d}\omega )=\exp
\left\{ -{\frac 12}\int_0^t|g(r)|^2\mathrm{d}\,r\right\}
\end{equation*}
of the standard Wiener process $w_t$. However, the different Brownian
motions $\widehat{w}_t(f)$ and $\widehat{w}_t(g)$ with $f^{*}g\neq g^{*}f$
do not have any classical representation on a single probability space $%
(\Omega ,\mathcal{F},\mathbf{P})$ because of noncommutativity (see A3 in
Appendix 3): 
\begin{equation*}
[\widehat{w}_t(f),\widehat{w}_t(g)]=\int_0^t\biggl(\bar f(r)g(r)-\bar
g(r)f(r)\biggr) \mathrm{d}\,r\,.
\end{equation*}

\noindent\textbf{Definition 1.} Let $\{\mathcal{A}_t|t\in \mathbf{R}_{+}\}$
be an increasing set of involutive subalgebras $\mathcal{A}_t\subseteq 
\mathcal{A}_s$, $t\leq s$ of the operators $X_t\in \mathcal{B}(\mathcal{D})$
generated by operators $X\in \mathcal{A}$, $\widehat{w}_t(g)$, $g\in 
\mathcal{K}$, such that 
\begin{equation*}
X_t\in \mathcal{A}_t\Leftrightarrow [X_t,Y]=0,\qquad \forall \,Y\in \mathcal{%
B}(\mathcal{D}):[X_0,Y]=0=\big[W_t(g),Y\big],
\end{equation*}
where the operators $X_0\in \mathcal{A}_0$, $W_t(g)$, $g\in L^2(0,t]$ are
assumed to act in $\mathcal{H}=\mathrm{H}\times \mathcal{F}$ as $X_0\otimes
\hat 1$ and $I\otimes \widehat{w}_t(g)$. A measurable operator function $%
F(t):\mathcal{D}\to \mathcal{D}$ is called adapted if $F(t)\in \mathcal{A}_t$
for almost all $t\in \mathbf{R}_{+}$.

We shall consider here only quantum stochastic integrals of the form 
\begin{equation}  \label{eq:diff1.1}
\imath _0^t(F,D)=\int_0^t\big(F(r)\,\mathrm{d}A_r+D(r)\,\mathrm{d}A_r^{*}%
\big),
\end{equation}
where $A_r=I\otimes \widehat{a}_r$, $A_r^{*}=I\otimes \widehat{a}_r^{*}$ and 
$F,D$ are locally square-integrable (see Appendix 4) together with adjoint $%
F^{*},D^{*}$ adapted operator-functions $\mathbf{R}_{+}\to \mathcal{A}_t$.
Note that on the exponential vectors, described by the product-functions $%
h(\tau )=k^{\otimes }(\tau )\,\psi $, where $\psi \in \mathrm{D}$ and $k\in
L^2(\mathbf{R}_{+}),$ the itegrals (\ref{eq:diff1.1}) are weakly defined as
the usual operator integrals 
\begin{equation*}
\big(h\mid \,\imath _0^t(F,D)\,h\big)=\int_0^t\Big(h\mid \,\big[ %
F(r)\,k(r)+D(r)\,\bar k(r)\big]\,h\Big)\,\mathrm{d}\,r.
\end{equation*}
This gives in particular, $\mathbf{E}[\imath _0^t(F,D)]=0$ for $\mathbf{E}%
[X]=(e|Xe)$, where $e(\tau )=\delta _\emptyset (\tau )\,\psi $ is the
exponential vector, corresponding to $k=0$. Such operator integrals on the
exponential domain were constructed by Hudson and Parthasarathy \cite%
{bib:diff12} for the case of bounded $F\left( t\right) ,D(t)$ .

For the adapted integrals (\ref{eq:diff1.1}), the quantum Ito formula \cite%
{bib:diff10}--\cite{bib:diff12} can be obtained as in the case of (\ref%
{eq:diffA.5}) (see Appendix 5), corresponding to $X_t(t)=0$ for almost all $%
t $. This formula defines (see Appendix 5) the pointwise multiplication 
\begin{equation*}
X(t)\,Y(t)=X(0)\,Y(0)+\int_0^t\mathrm{d}(XY)\,(r)
\end{equation*}
of the adapted operator-functions 
\begin{equation*}
X(t)=X(0)+\int_0^t\mathrm{d}X\,(r),\qquad Y(t)=Y(0)+\int_0^t\mathrm{d}Y\,(r)
\end{equation*}
in terms of the product $\mathrm{d}X(t)\,\mathrm{d}Y(t)=D(t)^{*}F(t)\,%
\mathrm{d}\,t$ of their stochastic differentials $\mathrm{d}X=D^{*}\mathrm{d}%
A+D\,\mathrm{d}A^{*}$, $\mathrm{d}Y=F^{*}\mathrm{d}A+F\,\mathrm{d}A^{*}$: 
\begin{eqnarray}
\mathrm{d}(XY)=&\mathrm{d}XY+X\mathrm{d}Y+\mathrm{d}X\mathrm{d}Y=D^*F\,%
\mathrm{d}\, t+(D^*Y+XF^*)\,\mathrm{d}A  \notag \\
&+(DY+XF)\,\mathrm{d}A^*.  \label{eq:diff1.2}
\end{eqnarray}

The classical Ito formula for the stochastic integrals%
\begin{equation*}
I_{0}^{t}(f,\omega )=\int_{0}^{t}f(r,\omega )\,\mathrm{d}w_{r}
\end{equation*}%
with respect to the standard Wiener process $w=\{w_{t}|t\in \mathbf{R}_{+}\}$
can be obtained from (\ref{eq:diff1.2}) by the Segal one-to-one
transformation $\omega :\widehat{w}_{t}\mapsto w_{t}$, where $\widehat{w}%
_{t}=\widehat{a}_{t}+\widehat{a}_{t}^{\ast }$. The latter represents the
adapted operator integrals $\imath _{0}^{t}(\widehat{f},\widehat{f}%
)=\int_{0}^{t}\widehat{f}(r)\,\mathrm{d}\,\widehat{w}_{r}$ for the
non-anticipated functionals $\widehat{f}(t)=f(t,\widehat{w})$ of commuting
selfadjoint operators $\widehat{w}=\{\widehat{w}_{t}\}$ with%
\begin{equation*}
\Vert \widehat{f}\delta _{\emptyset }\Vert _{t}^{2}:=\int_{0}^{t}\Vert 
\widehat{f}(r)\,\delta _{\emptyset }\Vert ^{2}<\infty
\end{equation*}%
in the form of the Ito integrals $\omega \lbrack {\iota }_{0}^{t}(\widehat{f}%
,\widehat{f})]=I_{0}^{t}(f,\omega )$, so that 
\begin{equation*}
\big\|\imath _{0}^{t}(\widehat{f},\widehat{f})\,\delta _{\emptyset }\big\|%
^{2}=\Vert \widehat{f}\,\delta _{\emptyset }\Vert _{t}^{2}=\int \big\|%
I_{0}^{t}(f,\omega )\big\|^{2}\mathbf{P}(\mathrm{d}\,\omega ),
\end{equation*}%
where $\mathbf{P}$ is the standard Wiener probability measure.

\noindent 1.2. \textbf{Quantum diffusion.} Quantum stochastic evolution in
the open system $\left\{ \mathcal{A}_t\right\} $ is described by an adapted
family $\{\iota (t)|t\in \mathbf{R}_{+}\}$ of $*$-representations $\iota
(t):X\mapsto X\left( t\right) $ of the initial algebra $\mathcal{A}\subseteq 
\mathcal{B}(\mathrm{D})$ into $\mathcal{B}(\mathcal{D})$ , i.e. of linear
maps $\mathcal{A}\to \mathcal{A}_t$, with the properties: 
\begin{equation*}
\iota (t,X^{*}X)=\iota (t,X)^{*}\iota (t,X),\qquad \iota (t,I)=I_0:=I\otimes
\hat 1.
\end{equation*}
It is called diffusion motion if the operator-valued functions $t\mapsto
X(t) $ have the quantum stochastic differentials of the form 
\begin{equation}  \label{eq:diff1.3}
\mathrm{d}X(t)+C(t)\,\mathrm{d}\,t=D^{*}(t)\,\mathrm{d}A_t+D(t)\,\mathrm{d}%
A_t^{*}.
\end{equation}
Here $C(t)=\gamma (t,X)$ is an adapted operator-valued function, locally
integrable ($p=1$) for every $X\in \mathcal{A}$ defined by the linear maps $%
\gamma (t):\mathcal{A}\to \mathcal{A}_t$, $t\in \mathbf{R}_{+}$ . $%
D^{*}(t)=\delta ^{*}(t,X)$, $D(t)=\delta (t,X)$ are adapted operator-valued
functions, locally square-integrable for each $X\in \mathcal{A}$, defined by
the linear maps $\delta ^{*}(t)$, $\delta (t):\mathcal{A}\to \mathcal{A}_t$.

Define the output process $Y=\{Y(t)|t\in \mathbf{R}_{+}\}$, which is subject
to measurement and described by a commutative family of (essentially)
selfadjoint operators $Y(t)=Y(t)^{*}$ on $\mathcal{D}$ with the initial
condition $Y(0)=0$ and stochastic differentials 
\begin{equation}  \label{eq:diff1.4}
\mathrm{d}Y(t)=G(t)\,\mathrm{d}\,t+F^{*}(t)\,\mathrm{d}A_t+F(t)\,\mathrm{d}%
A_t^{*}.
\end{equation}
Here $G(t)=G(t)^{*}\in \mathcal{A}_t$ is essentially self-adjoint and
locally-integrable ($p=1$). $F(t)\in \mathcal{A}_t$ is locally
square-integrable together with its conjugate: $F^{*}(t)=F(t)^{*}$ ; $G(t)$
and $F(t)$ are adapted operator-valued functions of $t\in \mathbf{R}_{+}$.

Unlike the classical case, not every involutive subalgebra $\mathcal{B}$ of $%
\mathcal{A}$ , but only a central one $\mathcal{B}\subset \mathcal{A}\cap 
\mathcal{A}^{\prime }$ , defines the conditional expectations $\mathbf{E}[X|%
\mathcal{B}]$ for any state vector $e\in \mathrm{H}$ as the positive
projections $\mathcal{A}\to \mathcal{B}$ which are compatible with $\mathbf{E%
}[X]=(e|Xe)$ such that $\mathbf{E}[\mathbf{E}[X|\mathcal{B}]]=\mathbf{E}[X]$
for all $X\in \mathcal{A}$. Hence, not every stochastic process described by
the equation (\ref{eq:diff1.4}), can be considered as an output process for
quantum diffusion, defined by equation (\ref{eq:diff1.3}), but only that for
which the posterior expectations of $X(t)$ with respect to the observation $%
Y(s)$, $s\leq t$, exist.

\noindent\textbf{Definition 2.} A process $Y(t)$ is called \emph{causal, or
nondemolition\/} with respect to the process $X(t)$ if 
\begin{equation}  \label{eq:diff1.5}
\big[X(t),Y(s)\big]:=X(t)\,Y(s)-Y(s)\,X(t)=0
\end{equation}
for all $t\geq s$, $s\in \mathbf{R}_{+}$.

The nondemolition condition together with the self-nondemolition of $Y$,
i.e. with the commutativity $[Y(t),Y(s)]=0$, $\forall \,t,s$, is necessary
and sufficient \cite{bib:diff11} for the existence of the conditional
expectations $\widehat{\pi }(t,X)=\mathbf{E}[X(t)|\mathcal{B}_{t}]$ for the
operators $X(t)=\iota (t,X)$ , with respect to the $\ast $-algebras 
\begin{equation*}
\mathcal{B}_{t}=\big\{Y\in \mathcal{B}(\mathcal{D})\mid \,[X,Y]=0,\qquad
\forall \,X\in \mathcal{B}(\mathcal{D}):\big[X,Y(s)\big]=0,\ \forall \,s\leq
t\big\}
\end{equation*}%
generated by the family $\{Y(s)|s\leq t\}$ and for every initial
vector-function $e\in \mathcal{D}$, $\Vert e\Vert =1$. If the process $Y$
with $Y\left( 0\right) =0$ is nondemolition with respect to the coefficients 
$C,D^{\ast },D$ of the equation (\ref{eq:diff1.3}), then it is nondemolition
with respect to the solution $X$ , corresponding to any initial $X\left(
0\right) \in \mathcal{A}$ . This and other sufficient conditions of the next
proposition obviously follow from the integral representation 
\begin{equation*}
X(t)=X-\int_{0}^{t}\big(C(r)\,\mathrm{d}\,r-D^{\ast }(r)\,\mathrm{d}%
A_{r}-D(r)\,\mathrm{d}A_{r}^{\ast }\big).
\end{equation*}

\noindent \textbf{Proposition 1.} The integrals $X\left( t\right) $ of (\ref%
{eq:diff1.3}) with $X\in \mathcal{A}$ are $\ast $-representations $\iota
(t):X\mapsto X(t)$ iff the linear maps 
\begin{equation*}
\gamma (t):\,X\longmapsto C(t),\qquad \delta ^{\ast }(t):\,X\longmapsto
D^{\ast }(t),\qquad \delta (t):\,X\longmapsto D(t)
\end{equation*}%
satisfy the following differential conditions

\begin{itemize}
\item[\textrm{(i)}] $\ \gamma (t,X^{*})=\gamma (t,X)^{*},\qquad \delta
(t,X^{*})=\delta ^{*}(t,X)^{*},\qquad \forall \,X\in \mathcal{A},$

\item[\textrm{(ii)}] $\gamma (t,X^{*}X)=\iota (t,X)^{*}\gamma (t,X)+\gamma
(t,X)^{*}\iota (t,X)-\delta (t,X)^{*}\delta (t,X),\\[0.5\baselineskip]
\delta (t,X^{*}X)=\iota (t,X)^{*}\delta (t,X)+\delta (t,X^{*})\,\iota
(t,X)=\delta ^{*}(t,X^{*}X)^{*},$

\item[\textrm{(iii)}] $\gamma (t,I)=0,\qquad \delta (t,I)=0=\delta
^{*}(t,I),\quad \forall \,t\in \mathbf{R}_{+}$.
\end{itemize}

The process $X(t)$ satisfies the condition (\ref{eq:diff1.5}) iff the
stochastic derivations $C(t)$,$D^{\ast }(t)$,$D(t)$ also satisfy the
condition (\ref{eq:diff1.5}) as $X(t)$ with respect to the nondemolition
process $Y(t)$ for all $X\in \mathcal{A}$, and the derivatives $G,\ F^{\ast
},\ F$ in (\ref{eq:diff1.4}) satisfy the differential nondemolition
conditions 
\begin{eqnarray}
\big[X(t),F^{\ast }(t)\big]=0 &=&\big[F(t),X(t)\big],\qquad \forall \,t\in 
\mathbf{R}_{+},  \notag \\
D^{\ast }(t)\,F(t)-F^{\ast }(t)\,D(t) &=&\big[G(t),X(t)\big].
\label{eq:diff1.6}
\end{eqnarray}

\textsc{Proof.} The stochastic differentials $\mathrm{d}X(t)=X(t+\mathrm{d}%
\,t)-X(t)$ of the linear $*$-maps $\iota (t):X\mapsto X(t)$ are defined by
the linear $* $-maps $\gamma (t),\ \delta ^{*}(t),\ \delta (t)$ by virtue of
linear independence of the fundamental differentials $\mathrm{d}\,t,\ 
\mathrm{d}A_t$ and $\mathrm{d}A_t^{*}$. The conditions (ii) are found by
applying the Ito formula (\ref{eq:diff1.2}) to $X(t)^{*}X(t)$: 
\begin{eqnarray*}
\mathrm{d}\,\big(X(t)^*X(t)\big)&=&\mathrm{d}X(t)^*\mathrm{d}X(t)+\mathrm{d}%
X(t)^*X(t)+X(t)^*\mathrm{d}X(t) \\
&=&\big[\delta(t,X)^*\delta(t,X)-\gamma(t,X)^*\iota(t,X)-\iota(t,X)^*%
\gamma(t,X)\big]\,\mathrm{d}\, t \\
&+&\mathrm{d}\,\imath_0^t \big(\delta(X)^*\iota(X)+\iota(X)^*\delta^*(X),\
\delta(X^*)\,\iota(X)+\iota(X)^*\delta(X)\big).
\end{eqnarray*}
By equating the stochastic derivatives of this differential and 
\begin{equation*}
\mathrm{d}\,\iota (t,X^{*}X)=\mathrm{d}\,\imath _0^t\big(\delta
^{*}(X^{*}X),\delta (X^{*}X)\big)-\gamma (X^{*}X)\,\mathrm{d}\,t,
\end{equation*}
we obtain that $\delta ^{*}$ and $\delta $ are the derivations of the
algebra $\mathcal{A}$, and $-\gamma $ has the positive-definite dissipator 
\begin{equation*}
\iota (X)^{*}\gamma (X)+\gamma (X)^{*}\iota (X)-\gamma (X^{*}X)=\delta
(X)^{*}\delta (X).
\end{equation*}
The condition (iii) follows from $\mathrm{d}\,\iota (t,I)=0$ because of the
independence of $\gamma ,\ \delta ^{*},\ \delta $.

If $Y(t)$ is a nondemolition process for $X(t)$, then 
\begin{equation*}
\big[\mathrm{d}X(t),Y(s)\big]=\big[X(t+\mathrm{d}\,t),Y(s)\big]-\big[%
X(t),Y(s)\big]=0
\end{equation*}%
with $t\geq s$ ; hence the nondemolition for $C,\ D^{\ast },\ D$: 
\begin{equation*}
\big[C(t),Y(s)\big]=0,\qquad \big[D^{\ast }(t),Y(s)\big]=0,\qquad \big[%
D(t),Y(s)\big]=0,\quad \forall \,t\geq s,
\end{equation*}%
follows by commutativity of $Y(s)$ with the indepedent differentials $%
\mathrm{d}\,t$,$\ \mathrm{d}A_{t}$ and$\ \mathrm{d}A_{t}^{\ast }$. Applying
equation (\ref{eq:diff1.2}) to the differential of the commutator $[X(t),\
Y(t)]=0$ we obtain (taking into account the equality $[\mathrm{d}X(t),\
Y(t)]=0$): 
\begin{eqnarray*}
\mathrm{d}\,\big[X(t),Y(t)\big] &=&\big[\mathrm{d}X(t),\mathrm{d}Y(t)\big]+%
\big[\mathrm{d}X(t),\mathrm{d}Y(t)\big]+\big[X(t),\mathrm{d}Y(t)\big] \\
&=&\Big(D^{\ast }(t)\,F(t)-F^{\ast }(t)\,D(t)+\big[X(t),G(t)\big]\Big)\,%
\mathrm{d}\,t \\
&&+\mathrm{d}\,\imath _{0}^{t}\big([X,F^{\ast }],[X,F]\big)=0,
\end{eqnarray*}%
which yields the differential self-nondemolition conditions (\ref{eq:diff1.6}%
). Hence, all conditions of the proposition are necessary.

\noindent\textbf{1.3. The Markov case. } The quantum diffusion (\ref%
{eq:diff1.3}) with coefficients 
\begin{equation*}
C(t)=\iota (t,C_t),\qquad D^{*}(t)=\iota (t,D_t^{*}),\qquad D(t)=\iota
(t,D_t),
\end{equation*}
corresponds to the Markov stochastic evolution (in strong sense). Here $%
C_t,D_t^{*},D_t\in \mathcal{A}$ are defined by the structural maps 
\begin{equation*}
\gamma _t\,:\,X\mapsto C_t,\qquad \delta _t^{*}\,:\,X\mapsto D_t^{*},\qquad
\delta _t\,:\,X\mapsto D_t,
\end{equation*}
for which the conditions (i)--(iii) indicated above were obtained by Hudson
and Evans in \cite{bib:diff14}. The self-nondemolition conditions (\ref%
{eq:diff1.6}) give the restrictions for the coefficients $G$ and $F^{*},F$
in this case. We shall restrict ourselves to consideration of the standard
case $F(t)=I_0=F^{*}(t)$ of the indirect measurement 
\begin{equation}  \label{eq:diff1.7}
Y(t)=\int_0^t\iota (t,G_r)\,\mathrm{d}\,r+I\otimes \widehat{w}_t
\end{equation}
of the diffusion of a square-integrable initial process defined locally by $%
G_t\in \mathcal{A}$ over the standard Wiener process $w_t$ represented in $%
\mathcal{H}$ by the operators $I\otimes \widehat{w}_t=A_t+A_t^{*}$. It is
not hard to prove \cite{bib:diff5} that $Y(r)=V_t(I\otimes \widehat{w}%
_r)\,V_t^{*}$, $\forall \,t>r\in \mathbf{R}_{+}$ by the uniqueness of the
stochastic operator equation 
\begin{equation*}
\mathrm{d}V_t+{\frac 18}\,G(t)\,V_t\mathrm{d}\,t=i\,G(t)\,V_t\mathrm{d}\,%
\widehat{u}_t,\qquad V_0=I,
\end{equation*}
where 
\begin{equation*}
G(t)=\iota (t,G_t),\qquad \widehat{u}_t={\frac i2}\,(\widehat{a}_t^{*}-%
\widehat{a}_t)={\frac 12}\,\widehat{w}_t(i).
\end{equation*}
The above implies the local unitary equivalence of the processes $%
A_t+A_t^{*} $ and $Y(t)$, which is always the case for the locally
norm-square-integrable operator-functions $G_t:\mathrm{H}\to \mathrm{H}$ 
\cite{bib:diff9}.

\noindent \textbf{Corollary 1.} In the case under consideration the
condition (\ref{eq:diff1.6}) completely defines the structure of the inner
derivation $X\mapsto [X,G_t]$ for $\delta _t-\delta _t^{*}$: 
\begin{equation}  \label{eq:diff1.8}
\delta _t(X)={\ \frac 12}\,[X,G_t]-\alpha _t(X),
\end{equation}
where $\alpha _t(X^{*})=\alpha _t(X)^{*}$, $\forall \,X\in \mathcal{A}$ is a 
$*$-derivation of the algebra $\mathcal{A}$. The conditions (i)--(iii) here
also define the structure of the maps $\gamma _t:\mathcal{A}\to \mathcal{A}$
in the form 
\begin{eqnarray}
\gamma_t(X)&=&{\frac{1}{2}}\,\lambda_t(X)-\beta_t(X),  \notag \\
\lambda_t(X)&=&{\frac{1}{4}}\,\big[G_t,[G_t,X]\big]+G_t\alpha_t(X)+%
\alpha_t(X)\,G_t -\alpha_t^2(X),  \label{eq:diff1.9}
\end{eqnarray}
where $\alpha _t^2(X)=\alpha _t(\alpha _t(X))$, $\beta _t:\mathcal{A}\to 
\mathcal{A}$ is some $*$-derivation 
\begin{equation*}
\beta _t(X^{*}X)=X^{*}\beta _t(X)+\beta _t(X^{*})\,X,\qquad \forall \,X\in 
\mathcal{A}.
\end{equation*}

\textsc{Proof.\/} By taking into account the differentiation property 
\begin{equation*}
[G_t,X^{*}X]=X^{*}[G_t,X]+[G_t,X^{*}]\,X,
\end{equation*}
and similarly for $\alpha _t(X^{*}X)$ we obtain 
\begin{eqnarray*}
\lambda_t(X^*X)&=&{\frac{1}{4}}\big[G_t,X^*[G_t,X]+[G_t,X^*]X\big] \\
&&+G_t\big(X^*\alpha_t(X)+\alpha_t(X^*)\,X\big)+\big(X^*\alpha_t(X)+
\alpha_t(X^*)\,X\big)\,G_t \\
&&-\big(X^*\alpha_t^2(X)+2\alpha_t(X)^*\alpha_t(X)+\alpha_t^2(X)^*X\big) \\
&=&X^*\lambda_t(X)+\lambda_t(X)^*X-{\frac{1}{2}}\,[X,G_t]^*[X,G_t] \\
&&+[X,G_t]^*\alpha_t(X)+\alpha_t(X)^*[X,G_t]-2\alpha_t(X)^*\alpha_t(X) \\
&=&X^*\lambda_t(X)+\lambda_t(X)^*X-2\delta_t(X)^*\delta_t(X).
\end{eqnarray*}
Hence $\gamma _t^0={\frac 12}\lambda _t$ possesses the property (ii) of the
map $\gamma _t$: 
\begin{equation*}
\gamma _t^0(X^{*}X)=X^{*}\gamma _t^0(X)+\gamma _t^0(X)^{*}X-\delta
_t(X)^{*}\delta _t(X),
\end{equation*}
and the $*$-property $\gamma _t^0(X^{*})=\gamma _t^0(X)^{*}$ , as does $%
\gamma _t$ in (iii). From here the result that $\beta _t=\gamma _t^0-\gamma
_t$ is a $*$-derivation follows.

The maps $\gamma _t$ are called generators for the Lindblad equation 
\begin{equation*}
\mathrm{d}\mu _0^t/\mathrm{d}t+\mu _0^t\circ \gamma _t=0,
\end{equation*}
where $\mu \circ \gamma (X)=\mu (\gamma (X))$, which is an algebraic analog
of the Kolmogorov equation. This is satisfied by the operators $X_0^t=\mu
_0^t(X)$, $X\in \mathcal{A}$ of the conditional expectation $\mu
_0^t(X)=(\delta _\emptyset |X(t)\,\delta _\emptyset )$ , with respect to the
vacuum function $\delta _\emptyset \in \mathcal{F}$ defined by 
\begin{equation*}
X_0^t\psi =\big[\iota (t,X)\,h\big]\,(\emptyset ),\qquad \forall \,h=\psi
\otimes \delta _\emptyset ,\quad \psi \in \mathrm{D}.
\end{equation*}

The differential conditions obtained for Markov diffusion are necessary for
the existence of a unique solution $X(t)$ of equation (\ref{eq:diff1.3}) for
all $X(0)=X\in \mathcal{A}$ and so, for the Lindblad equation. They are
sufficient in the case \cite{bib:diff14} of constant $\alpha _t,\ \beta _t$
and $G_t$ and the boundedness of the algebra $\mathcal{A}$ (for instance,
with $\mathrm{D}=\mathrm{H}$). Moreover, as was proved in \cite{bib:diff10},
the maps $\iota (t):X\mapsto X(t)$ are representations of $\mathcal{A}$ in $%
\mathcal{A}_t$, satisfying the condition of nondemolition (\ref{eq:diff1.5})
for all $X\in \mathcal{A}$. This is also true under significantly more
general conditions of local $p$-integrability over the norm of the operators 
$G_t:\mathrm{H}\to \mathrm{H}$ (with $p=2$) and the maps $\alpha _t\ (p=2)$
and $\beta _t\ (p=1)$ from $\mathcal{A}\subseteq \mathcal{B}(\mathrm{H})$ to 
$\mathcal{A}$.

In the case of inner derivations 
\begin{equation*}
\alpha _t(X)=i[S_t,X],\qquad \beta _t(X)=i\Big[H_t+{\frac 14}%
\,(S_tG_t+G_tS_t),X\Big],
\end{equation*}
(as is the case for the von-Neumann algebra $\mathcal{A}$), a quantum Markov
diffusion is defined by structure maps of the type 
\begin{eqnarray}
\delta_t(X)=[X,L_t],& &\delta_t^*(X)=[L_t^*,X],  \notag \\
\gamma_t(X)={\frac{1}{2}}\,\big(L_t^*[L_t,X]&+&[X,L_t^*]\,L_t\big)+i[X,H_t],
\label{eq:diff1.10}
\end{eqnarray}
where 
\begin{equation*}
L_t={\frac 12}\,G_t+iS_t,\qquad L_t^{*}={\frac 12}\,G_t-iS_t,\qquad
H_t=H_t^{*}\in \mathcal{A}_t.
\end{equation*}

\section{Quantum diffusion and filtering.}

\setcounter{equation}{0}

\typeout{Quantum Diffusion and Filtering.}

\noindent \textbf{2.1. The a posteriori dynamics.} The quantum diffusion (%
\ref{eq:diff1.3}) under nondemolition measurement (\ref{eq:diff1.4}), is
described by the classical random variables $x_t(\omega )=\langle
X(t)\rangle _t(\omega )$ of the conditional expectations $\langle
X(t)\rangle _t=\mathbf{E}[X(t)|\mathcal{B}_t]$ , for the operators $%
X(t)=\iota (t,X)$ , on the trajectories $\omega \in \Omega $ of the process $%
Y(t)$ , with respect to initial vector-valued function $e_0=\psi _0\otimes
\delta _\emptyset $, $\psi _0\in \mathrm{D}$. As was established for the
first time in \cite{bib:diff5}, \cite{bib:diff6}, the random process $%
x_t:\Omega \to \mathbf{C}$ , considered as a stochastic map $X(t)\mapsto
x_t(\omega )$, satisfies the Ito filtering equation for quantum diffusion 
\begin{equation}  \label{eq:diff2.1}
\mathrm{d}\,\big\langle \iota (t,X)\big\rangle_t+\big\langle \gamma (t,X)%
\big\rangle_t\mathrm{d}\,t=\big\langle \kappa (t,X)-\big\langle G(t)%
\big\rangle_t\iota (t,X)\big\rangle_t\mathrm{d}\,\widetilde{Y}
\end{equation}
with respect to the stochastic map $\langle \cdotp\rangle _t:X(t)\mapsto x_t$%
. In this equation $\widetilde{Y}(t)=Y(t)-\int_0^t\big\langle G(r)\big\rangle%
_r\mathrm{d}\,r$ is an innovation martingale for the observed process (\ref%
{eq:diff1.4}), and $\kappa (t):\mathcal{A}\to \mathcal{B}(\mathrm{D})$ is a
linear $*$-map, which in the case of $F(t)=I_0=F^{*}(t)$ is of the
particularly simple form: 
\begin{equation}  \label{eq:diff2.2}
\kappa (t,X)={\frac 12}\,\big(G(t)\,X(t)+X(t)\,G(t)\big)-\alpha (t,X),
\end{equation}
where $\alpha (t,X)$ is defined by the $*$-derivation $\delta (t)+\delta
^{*}(t)=-2\alpha (t)$. Equation (\ref{eq:diff2.1}), derived in \cite%
{bib:diff5} by means of the martingale methods of quantum nonlinear
filtering, extends the basic equation (8.10), \cite{bib:diff18} of the
optimal diffusion filtering to the case of the noncommutative operator
algebras $\mathcal{A}$. By complete analogy with the classical case the
quantum Ito formula was used with the innovation process and the
representation theorem \cite{bib:diff11}, which requires the conditional
expectations $x_t(\omega )$ to exist. This requirement is met by the
self-nondemolition condition (\ref{eq:diff1.5}) for all $s\leq t$ which is
trivially sutisfied in the commutative case $s>t$.

In the quantum Markov case of the indirect measurement (\ref{eq:diff1.7})
the conditional expectation $x_t(\omega )=\pi _t(X,\omega )$ of the
operators $X(t)$ can be found as in the classical case by solving an
autonomous stochastic equation for the a posteriori state $\widehat{\pi }%
_t(X)=\langle \iota (t,X)\rangle _t$. The latter is defined on the
trajectories $\omega $ as a linear stochastic positive normalized map $%
\omega (\widehat{\pi }_t(X))=x_t(\omega )$ of the algebra $\mathcal{A}$ into 
$\mathbf{C}$ satisfying the condition 
\begin{equation*}
\int x_t(\omega )\,y(\omega )\,\mathbf{P}_0^t(\mathrm{d}\,\omega )=\big(%
e_0\mid X(t)\,Ye_0\big),\qquad \forall \,X\in \mathcal{A}.
\end{equation*}
In this equation, $Y\in \mathcal{B}_t$ is any bounded operator in the
algebra of the observed $\mathcal{B}_t$, $y(\omega )=\omega (\widehat{y})$
is the Segal transformation of the operator $\widehat{y}$ in $\mathcal{F}$
that corresponds to the unitary-equivalent operator $I\otimes \widehat{y}%
=V_t^{*}YV_t$, and $\mathbf{P}_0^t$ is an induced (by the unitary
transformation) probability measure on the trajectories $\omega
|[0,t):=\{w_r|r\in [0,t)\}$ , restricted to the interval $[0,t)$ with
respect to the initial vector-state 
\begin{equation*}
\int y(\omega )\,\mathbf{P}_0^t({\mathrm{d}\,}\omega )=\big(V_t^{*}(\psi
_0\otimes \delta _\emptyset )\big)\,\mid \,(I\otimes \widehat{y}%
)\,V_t^{*}(\psi _0\otimes \delta _\emptyset )\big).
\end{equation*}

\noindent\textbf{2.2. The filtering equation.} Let us sketch the essentials
in the derivation of a stochastic Markov quantum filtering equation,
obtained for the general output process in \cite{bib:diff9}. First, we shall
prove that the vacuum conditional expectation $\mu _g^t(X)=(\delta
_\emptyset |\pi _g(t,X)\,\delta _\emptyset )$ of the product $\pi
_g(t,X)=\iota (t,X)e_g(t)$, where 
\begin{equation}  \label{eq:diff2.3}
e_g(t)=\exp \bigg\{\int_0^tg(r)\,\mathrm{d}Y(r)-{\frac 12}\,g(r)^2\mathrm{d}%
\,r\bigg\},
\end{equation}
satisfies the linear evolution equation 
\begin{equation}  \label{eq:diff2.4}
{\frac{\mathrm{d}\,}{\mathrm{d}\,t}}\,\mu _g^t(X)+\mu _g^t\circ \gamma
_t(X)=\mu _g^t\big(G_t\cdotp X-\alpha _t(X)\big)\,g(t),
\end{equation}
where $\mu _g^0(X)=X$, and $G\cdotp X=(GX+XG)/2$. Let us asume the
uniqueness of this solution, which is always true for locally $p$-integrable
bounded (over the norm) maps $G_t,\ \alpha _t$ $(p=2)$ and $\beta _t\ (p=1)$%
. We shall prove that $\mu _g^t(X)$ is the mathematical expectation of the
product $\widehat{\mu }_g^t(X)=\widehat{\mu }^t(X)\,\hat e_g^t$ , of the
stochastic operators $\mu ^t(X,\omega )=\omega [\widehat{\mu }^t(X)]$ (which
satisfy a quantum filtering equationt) with the exponentials 
\begin{equation*}
e_g^t(\omega )=\exp \int_0^t{\Big[}g(r)\,\mathrm{d}w_r-{\frac 12}\,g(r)^2%
\Big]\,\mathrm{d}\,r=\omega (\hat e_g^t),
\end{equation*}
(which are defined with respect to the trajectories $\omega :\,t\mapsto w_t$
of the standard Wiener process $w_t,\ t\in \mathbf{R}_{+}$). The above means
that the output process $Y(r)$, restricted by any $t\in \mathbf{R}_{+}$ , is
absolutely continuous with respect to the standard restricted process $%
w^t=\{w_r|r<t\}$. This follows from the unitary equivalence $%
Y(r)=V_tY_rV_t^{*},\forall r<t$ and $Y_r=I\otimes \widehat{w}_r,\ \forall
t\in \mathbf{R}_{+}$ , representing the output up to a time $t$ with respect
to the initial vector-function $e_0=\psi _0\otimes \delta _\emptyset $ and $%
e_0^t=V_t^{*}e_0$ correspondingly. The probability density $\rho _0^t(\omega
)=\mathbf{P}_0^t(\mathrm{d}\,\omega )/\mathbf{P}(\mathrm{d}\,\omega )$ , for
the measurement of the trajectory $\{w_r|r<t\}$ , of the process $Y$ on the
interval $[0,t)$ , is defined with respect to the standard Wiener
probability measure $\mathbf{P}(\mathrm{d}\,\omega )$ by the formula $\rho
_0^t(\omega )=\varphi _0^t(I,\omega )$. Here $\varphi _0^t(\omega )=\varphi
_0\circ \mu ^t(\omega )$ is the stochastic functional $\varphi _0^t(X,\omega
)=(\psi _0|\mu ^t(X,\omega )\,\psi _0)$, which corresponds to the initial
state $\varphi _0(X)=(\psi _0|X\psi _0)$ on the algebra $\mathcal{A}$.
Finally, we deduce a nonlinear equation for the a posteriori state $\pi
_t(\omega )$ using the ordinary Ito formula and the normalization of the
stochastic functional $\varphi _0^t(\omega )$.

\begin{theorem}
Let the equation (\ref{eq:diff2.4})has the unique solution $X_{g}^{t}=\mu
_{g}^{t}(X)$, corresponding to the initial condition $\mu _{g}^{0}(X)=X$ for
each $X\in \mathcal{A}$ and $g\in \mathcal{K}$ . Then it coincides with the
vacuum expectation $X_{g}^{t}=(\delta _{\emptyset }|\pi _{g}(t,X)\,\delta
_{\emptyset })$, where $\pi _{g}(t,X)=\iota (t,X)e_{g}(t)$ and is defined by
the Wiener average 
\begin{equation}
X_{g}^{t}=\int_{0}^{t}X^{t}(\omega )\,\exp \bigg\{\int_{0}^{t}g(r)\,\mathrm{d%
}w_{r}-{\frac{1}{2}}\,g(r)^{2}\mathrm{d}\,r\bigg\}\,\mathbf{P}(\mathrm{d}%
\,\omega )  \label{eq:diff2.5}
\end{equation}%
over the continuous trajectories $\omega \in \Omega $. Here $X^{t}(\omega
)=\omega \lbrack \widehat{\mu }^{t}]$ is the Segal transformation of the
solution $\widehat{X}^{t}=\widehat{\mu }^{t}(X)$ to the operator filtering
equation 
\begin{equation}
\mathrm{d}\,\widehat{\mu }^{t}(X)+\widehat{\mu }^{t}\circ \gamma _{t}(X)\,%
\mathrm{d}\,t=\widehat{\mu }^{t}\big(G_{t}\cdot X-\alpha _{t}(X)\big)\,%
\mathrm{d}\,\widehat{w}_{t}  \label{eq:diff2.6}
\end{equation}%
with initial condition $\widehat{\mu }^{0}(X)=X$. In this case the linear
stochastic equation (\ref{eq:diff2.6}) also has a unique solution in the Ito
sense, which defines almost everywhere ($\rho _{0}^{t}(\omega )\neq 0$) for
each $\psi _{0}\in \mathrm{H}$ the a posteriori state 
\begin{equation*}
\pi _{t}(X,\omega )={\frac{(\psi _{0}|X^{t}(\omega )\,\psi _{0})}{\rho
_{0}^{t}(\omega )}},
\end{equation*}%
where the probability density $\rho _{0}^{t}(\omega )=(\psi
_{0}|I^{t}(\omega )\,\psi _{0})$ is given by the positive operator $%
I^{t}(\omega )=\mu ^{t}(I,\omega )$, satisfying the martingale property%
\begin{equation*}
\int I^{t}(\omega )\mathbf{P}(\mathrm{d}\omega |w^{r})=I^{r}(\omega
),\forall r<t.
\end{equation*}
\end{theorem}

\textsc{Proof.} First we find a quantum stochastic equation for $X_g(t)=\pi
_g(t,X)$ using the Ito formula 
\begin{equation*}
\mathrm{d}e_g(t)=g(t)\,e_g(t)\,\mathrm{d}Y(t),\qquad e_g(0)=1,
\end{equation*}
where 
\begin{equation*}
\mathrm{d}Y(t)=G(t)\,\mathrm{d}\,t+F(t)^{*}\mathrm{d}A_t+\mathrm{d}%
A_t^{*}F(t).
\end{equation*}
We obtain according to (\ref{eq:diff1.2}) 
\begin{eqnarray*}
\mathrm{d}\,\big(X(t)\,e_g(t)\big)&=&\mathrm{d}X(t)\,\mathrm{d}\,\,e_g(t)+%
\mathrm{d}X(t)\,e_g(t)+X(t)\,\mathrm{d}\,\,e_g(t) \\
&=&(g\,D^*F-C+gXG)(t)\,e_g(t)\,\mathrm{d}\, t \\
&&+\mathrm{d}\,\imath_0^t(D^*+X\,F^*gD+X\,Fg)\,e_g(t) \\
&=&\pi_g\Big(t,(G_tX+X\,G_t)\,{\frac{g(t)}{2}}-\alpha_t(X)\,g(t)-\gamma_t(X)%
\Big)\,\mathrm{d}\, t \\
&&+\pi_g\big(t,\delta_t^*(X)+g(t)\,X\big)\,\mathrm{d}A_t +\pi_g\big(%
t,\delta_t(X)+g(t)\,X\big)\,\mathrm{d}A_t^*,
\end{eqnarray*}
where the explicit form (\ref{eq:diff1.8}) has been used for 
\begin{equation*}
D^{*}=\delta ^{*}(X),\qquad D=\delta (X),\qquad C=\gamma (X),\qquad GX+D=G%
\cdotp X-\alpha (X)
\end{equation*}
with $F=I_0=F^{*}$.

Taking into account the martingale property of the quantum stochastic
integral (\ref{eq:diff1.1}) with respect to the vacuum-vector $\delta
_\emptyset \in \mathcal{F}$, we find the equation (\ref{eq:diff2.4}) for the
operator 
\begin{equation*}
\mu _g^t(X):\psi \longmapsto \mu _g^t(X)\,\psi =[\pi _g(t,X)h](\emptyset )
\end{equation*}
in $\mathrm{H}$ is implied by the action of $\pi _g(t,X)=X(t)\,e_g(t)$ on $%
h=\psi \otimes \delta _\emptyset $: 
\begin{equation*}
\mathrm{d}\,\mu _g^t(X)=\mu _g^t\big((G_t\cdotp X-\alpha _t(X)\,g(t)-\gamma
_t(X)\big)\mathrm{d}t,\qquad \forall \,X\in \mathcal{A},
\end{equation*}
where $\gamma _t:\mathcal{A}\to \mathcal{A}$ is defined in the form (\ref%
{eq:diff1.10}).

Now, if $X^t(\omega )=\mu ^t(X,\omega )$ satisfies the stochastic equation 
\begin{equation*}
\mathrm{d}X^t(\omega )+C^t(\omega )\,\mathrm{d}\,t=D^t(\omega )\,\mathrm{d}\,%
\widehat{w}_t,\qquad X^0(\omega )=X,
\end{equation*}
we can derive a differential for 
\begin{equation*}
\widehat{\mu }_g^t(X)=\widehat{\mu }^t(X)\,\hat e_g^t,\qquad \hat e_g^t=\exp %
\bigg\{\int_0^t\Big(g(r)\,\mathrm{d}\,\widehat{w}_r-{\ \frac 12}\,g(r)^2%
\mathrm{d}\,r\Big)\bigg\},
\end{equation*}
by means of the Ito formula of classical stochastic calculus.%
\begin{equation*}
\mathrm{d}X^t(\omega )+C^t(\omega )\,\mathrm{d}\,t=D^t(\omega )\,\mathrm{d}\,%
\widehat{w}_t,\qquad X^0(\omega )=X.
\end{equation*}
Using $\mathrm{d}\,\hat e_g^t=g(t)\hat e_g^t\mathrm{d}\,\widehat{w}_t$ we
have 
\begin{eqnarray*}
\mathrm{d}(\widehat X^t\hat e_g^t)&=&\mathrm{d}\,\widehat X^t\mathrm{d}\,
\hat e_g^t+\mathrm{d}\,\widehat X^t\hat e_g^t+ \widehat X^t\mathrm{d}\,\hat
e_g^t \\
&=&\big(g(t)\,\widehat D^t-\widehat C^t\big)\, \hat e_g^t\mathrm{d}\, t+\big(%
\widehat D^t+\widehat X^tg(t)\big)\, \hat e_g^t\mathrm{d}\,\widehat w_t,
\end{eqnarray*}
what can be written in the form of the stochastic equation%
\begin{equation*}
\mathrm{d}X_g^t\left( \omega \right) +(C_g^t\left( \omega \right)
-D_g^t\left( \omega \right) g\left( t\right) )\mathrm{d}t=(D_g^t\left(
\omega \right) +X_g^t\left( \omega \right) g\left( t\right) )\mathrm{d}w_t
\end{equation*}
for $X_g^t\left( \omega \right) =\omega \left( \widehat{X}^t\widehat{e}%
_g^t\right) $. Hence, the mathematical expectation (\ref{eq:diff2.5}) of $%
X_g^t(\omega )=X^t(\omega )e_g^t(\omega )$ with respect to the Gaussian
measure $\mathbf{P}$ of the standard Wiener process $w$ satisfies the
equation 
\begin{equation*}
\mathrm{d}\,\mu _g^t(X)=\big(D_g^tg(t)-C_g^t\big)\,\mathrm{d}\,t,\qquad \mu
_g^0(X)=X.
\end{equation*}
Comparison of this equation with equation (\ref{eq:diff2.4}) gives the
coefficients 
\begin{equation*}
C_g^t=\int C^t(\omega )\,e_g^t(\omega )\,\mathbf{P}(\mathrm{d}\,\omega
),\qquad D_g^t=\int D^t(\omega )\,e_g^t(\omega )\,\mathbf{P}(\mathrm{d}%
\,\omega )
\end{equation*}
in the form: 
\begin{equation*}
C_g^t=\mu _g^t\big(\gamma _t(X)\big),\qquad D_g^t=\mu _g^t\big(G_t\cdotp %
X-\alpha _t(X)\big).
\end{equation*}
Consequently $C^t(\omega )=\omega (\widehat{C}^t)$, $D^t(\omega )=\omega (%
\widehat{D}^t)$ are the coefficients 
\begin{equation*}
\widehat{C}^t=\widehat{\mu }^t\big(\gamma _t(X)\big),\qquad \widehat{D}^t=%
\widehat{\mu }^t\big(G_t\cdotp X-\alpha _t(X)\big),
\end{equation*}
that define an equation for $X^t(\omega )=\mu ^t(X,\omega )$ in the form of (%
\ref{eq:diff2.6}). The solution $I^t(\omega )=\mu ^t(I,\omega )$ to this
equation for the initial condition $X=I$ defines a positive operator-valued
diffusive process $I^t(\omega )=\omega (\widehat{\mu }^t(I))$ which
satisfies to the martingale equation $\mathrm{d}I^t\left( \omega \right)
=G^t\left( \omega \right) \mathrm{d}w_t$ with the initial condition $%
I^0\left( \omega \right) =I$ , where $G^t\left( \omega \right) =\mu ^t\left(
G,\omega \right) $ and the properties $\alpha _t\left( I\right) =0=\gamma
_t\left( I\right) $ are substituted into (\ref{eq:diff2.6}). Thus the
Theorem 1 is proved.

\noindent \textbf{2.3. The classical case.} The remark that follows provides
an explanation why equation (\ref{eq:diff2.6}) is a noncommutative analog of
the Zakai filtering equation.

\noindent \textbf{Remark 1.} Let $\mathcal{A}$ be a commutative algebra
equivalent to the space $C^\infty (\mathbf{R}^d)$ of
infinitely-differentiable functions $x:\mathbf{R}^d\to \mathbf{C}$ with the
pointwise product, and the involution $x^{*}(z)=\bar x(z)$. Then equation (%
\ref{eq:diff2.6}) is an operator representation of the Zakai equation 
\begin{equation}  \label{eq:diff2.7}
\mathrm{d}\,\mu _{z_0}^t+{\frac 12}\,\Delta _t\mu _{z_0}^t\mathrm{d}%
\,t=(g_t+\nabla _t)\,\mu _{z_0}^t\mathrm{d}w_t,\qquad \mu _{z_0}^0=\delta
_{z_0},
\end{equation}
for the nonnormalized a posteriori distribution $\mu _{z_0}^t(\mathrm{d}%
z,\omega )$ of the Markov diffusion process $z(t)$ described by the
stochastic equation 
\begin{equation}  \label{eq:diff2.8}
\mathrm{d}z+c_t(z)\,\mathrm{d}\,t=a_t(z)\,\mathrm{d}v_t,\qquad z(0)=z_0,
\end{equation}
with the indirect measurement 
\begin{equation*}
\mathrm{d}y(t)=g_t(z(t))\mathrm{d}\,t+\mathrm{d}w_t
\end{equation*}
defined by the standard Wiener process $w_t=-v_t$. Here 
\begin{eqnarray*}
\int x(z)\,\nabla_t\mu(\mathrm{d}z)&=&-\int\sum_{k=1}^da^k(z)\,x_k^{%
\prime}(z)\,\mu(\mathrm{d}z), \\
\int x(z)\,\Delta_t\mu(\mathrm{d}z)&=&+2\int\sum_{k=1}^dc_t^k(z)\,x_k^{%
\prime}(z)\,\mu(\mathrm{d}z) \\
&&-\int\sum_{k,l=1}^da_t^k(z)\, a_t^l(z)\,x_{kl}^{\prime\prime}(z)\,\mu(%
\mathrm{d}z),
\end{eqnarray*}
with $x_k^{\prime }=\partial _kx$, $x_{kl}^{\prime \prime }=\partial
_k\partial _lx$. The integral $\int \mu _{z_0}^t(\mathrm{d}z,\omega )=\rho
_{z_0}^t(\omega )$ defines the probability density of the output process $%
y(r,\omega )$ on the interval $0\leq r<t$ with respect to the Wiener
distribution $\mathbf{P}(\mathrm{d}\,\omega )$ with given initial state $%
z_0\in \mathbf{R}^d$, $\delta _{z_0}(\mathrm{d}z)=1$ with $z_0\in \mathrm{d}%
z $, $\delta _{z_0}(\mathrm{d}z)=0$, and $z_0\notin \mathrm{d}z$.

Indeed, in the case of the commutative algebra $\mathcal{A}\simeq C^\infty (%
\mathbf{R}^d)$ , $G_t$ is the multiplication operator by the given function $%
g_t(z)$ of the state $z\in \mathbf{R}^d$ of the Markov process $z(t)$ . This
process has the generator $\gamma _t(X)(z)=[\Gamma _tx](z)$, defined by the
diffusion operator $\Gamma _t$ on the measurable functions $x:z\mapsto x(z)$%
, $x\in C^\infty (\mathbf{R}^d)$. The indirect measurement of $g_t(z)$ is
given by the output process 
\begin{equation*}
y(t)=\int_0^tg_r(z(r))\,\mathrm{d}\,r+w_t.
\end{equation*}
Since $[G_t,X]=0$, $\forall X\in \mathcal{A}$, $\delta _t=-\alpha _t$ is a
real derivation 
\begin{equation*}
\alpha _t(X)\,(z)=a_t(z)\,\partial x(z):=a_t^k(z)\,\partial _kx(z),\qquad
\partial _k={\frac \partial {\partial z^k}},
\end{equation*}
and $\gamma _t=G_t\cdotp\alpha _t-\beta _t-{\frac 12}\alpha _t^2$, where $G_t%
\cdotp\alpha _t(X)(z)=g_t(z)\alpha _t(X)(z)$, the operation 
\begin{equation}  \label{eq:diff2.9}
\Gamma _t=(a_tg-b_t)\,\partial -{\frac 12}\,(a_t\partial )^2=c_t^k\partial
_k-{\frac 12}\,a_t^ka_t^l\partial _k\partial _l
\end{equation}
is a standard generator of the diffusion process $z(t)$, with $%
c_t^k=a_t^l(g_t\delta _l^k-{\frac 12}\partial _la_t^k)-b_t^k$.

Note that the noise $v_t=-w_t$ in the classical system appeared essentially
the same as in the observation channel because it was represented in the
Fock space of the Wiener process $w_t$. In order to represent a classical
stochastic system in the same way with the noise $\upsilon _t\neq 0$ which
is independent of $w_t$, it is necessary to start from the Fock space $%
\mathcal{F}=\Gamma (\mathcal{K})$ over $\mathcal{K}=\mathbf{C}^m\otimes L^2(%
\mathbf{R}_{+})$ with multiplicity $m\geq 2$, as is the case in \cite%
{bib:diff14}.

\noindent \textbf{2.4. The a posteriori equation.} In the general case the
filtering equation (\ref{eq:diff2.6}) defines the nonnormalized a posteriori
state $\widehat{\varphi }_0^t=\varphi _0\circ \widehat{\mu }_0^t$, which is
the vector-state $\widehat{\varphi }_0^t(X)=(\widehat{\psi }_0^t|X\widehat{%
\psi }_0^t)$ for all $\varphi _0(X)=(\psi _0|X\psi _0)$ in the case of inner
derivations (\ref{eq:diff1.10}). The normalized a posteriori state $\widehat{%
\pi }_t(X)=\widehat{\varphi }_0^t(X)/\widehat{\varphi }_0^t(I)$ satisfies
the (nonlinear) a posteriori equation 
\begin{equation}  \label{eq:diff2.10}
\mathrm{d}\,\widehat{\pi }_t(X)+\widehat{\pi }_t\circ \gamma _t(X)\,\mathrm{d%
}\,t=\widehat{\pi }_t\big(\kappa _t(X)-\widehat{\pi }_t(G_t)\,X\big)\,%
\mathrm{d}\,\widetilde{w}_t,
\end{equation}
with initial condition $\widehat{\pi }_0(X)=\varphi _0(X)$ , where 
\begin{equation*}
\kappa _t(X)=G_t\cdotp X-\alpha _t\left( X\right) ,\mathrm{d}\,\widetilde{w}%
_t=\mathrm{d}w_t-\widehat{\pi }_t(G_t)\,\mathrm{d}\,t.
\end{equation*}
The nonlinear stochastic equation (\ref{eq:diff2.10}) is the Markov case of
the general a posteriori diffusive equation (\ref{eq:diff2.1}) with
innovating martingale $\mathrm{d}\tilde Y$ represented by $\mathrm{d}\tilde
w $. It can be deduced from the linear one (\ref{eq:diff2.6}) for the
nonnormalized state $\widehat{\varphi }_t(X)=\varphi _0(\widehat{{\mu }}%
_0^t(X))$ by applying the classical Ito formula to the product $\widehat{%
\varphi }_t(X)=\widehat{\rho }_0^t\widehat{\pi }_t(X)$ and noting that the
positive martingale $\widehat{\rho }_0^t=\widehat{\varphi }_0^t(I)$ has the
stochastic differential $\mathrm{d}\,\widehat{\rho }_0^t=\widehat{\varphi }%
_0^t(G_t)\mathrm{d}\,\widehat{w}_t$. Indeed, 
\begin{eqnarray*}
\mathrm{d}\,\widehat\varphi_0^t(X)&=&\widehat\rho_0^t\mathrm{d}%
\,\widehat\pi_t(X)+ \mathrm{d}\,\widehat\rho_0^t\widehat\pi_t(X)+\widehat%
\rho_0^t\mathrm{d}\,\widehat\pi_t(X) \\
&=&\widehat\varphi_0^t(G_t)\,\widehat\pi_t\circ\widetilde{\kappa}_t(X)\,%
\mathrm{d}\, t+ \big(\widehat\varphi_0^t(G_t)\,\widehat\pi_t(X)+
\widehat\rho_0^t\widehat\pi_t\circ\widetilde{\kappa}_t(X)\big)\,\mathrm{d}%
\,\widehat w_t \\
&&-\widehat\rho_0^t\big(\widehat\pi_t\circ\gamma_t(X)+ \widehat\pi_t\circ%
\widetilde{\kappa}_t(X)\,\widehat\pi_t(G_t)\big)\,\mathrm{d}\, t \\
&=&\varphi_0^t\big(G_t\cdotp X-\alpha_t(X)\big)\,\mathrm{d}\,\widehat w_t-
\widehat\varphi_0^t\circ\gamma_t(X)\,\mathrm{d}\, t,
\end{eqnarray*}
where $\widetilde{\kappa }_t(X)=\kappa _t(X)-\widehat{\pi }_t(G_t)X$. Note
that in the deduction of the equation the following relation was used: 
\begin{equation*}
\widehat{\rho }_0^t\widehat{\pi }_t\circ \widetilde{\kappa }_t(X)\,\widehat{%
\pi }_t(G_t)=\widehat{\pi }_t\circ \widetilde{\kappa }_t(X)\,\widehat{%
\varphi }_0^t(G_t).
\end{equation*}

\section{Linear quantum diffusion with observation}

\setcounter{equation}{0}

Let $\Xi$ be a symplectic $\sharp$-space, i.e.\ a complex space with the
involution 
\begin{equation*}
\eta\in\Xi\mapsto\eta^{\sharp},\qquad \eta^{\sharp\sharp}=\eta,\qquad \Big(%
\sum\lambda_i\eta_i\Big)^{\sharp}=\sum\lambda_i^*\eta_i^{\sharp},\quad
\forall\,\lambda_i\in\mathbf{C},
\end{equation*}
and skew-symmetric bilinear $\sharp$-form: $s:\Xi\times\Xi\to\mathbf{C}$, 
\begin{equation*}
s(\eta,\eta^\sharp)=-s(\eta^\sharp,\eta),\qquad s(\eta^\sharp,\eta)^*=
s(\eta,\eta^\sharp).
\end{equation*}

We denote by $\func{Re}\,\Xi $ the real space of the $\sharp $-invariant
vectors\break $\eta =\eta ^{\sharp }\in \Xi $, and assume that $\func{Re}%
\,\Xi $ is a Hilbert space with respect to the scalar product $\langle \xi
,\eta \rangle =\langle \eta ,\xi \rangle $, satisfying the inequality 
\begin{equation*}
\xi ^{2}\eta ^{2}-\langle \xi ,\eta \rangle ^{2}\geq {\frac{1}{4}}\,s(\xi
,\eta )^{2},\qquad \forall \,\xi ,\eta \in \func{Re}\,\Xi ,
\end{equation*}%
where $\xi ^{2}=\langle \xi ,\xi \rangle $, $\eta ^{2}=\langle \eta ,\eta
\rangle $.

A linear map $R:\Xi \rightarrow \mathcal{B}(\mathrm{D})$, satisfying the $%
\sharp $-property $R(\eta )^{\ast }=R(\eta ^{\sharp })$ defines an operator
representation of the canonical commutation relations if 
\begin{equation}
\big[R(\eta ),R(\eta ^{\sharp })\big]=R(\eta )\,R(\eta ^{\sharp })-R(\eta
^{\sharp })R(\eta )={\frac{1}{i}}\,s(\eta ,\eta ^{\sharp })\,I.
\label{eq:diff3.1}
\end{equation}%
on a complex pre-Hilbert space $\mathrm{D}$. It is called Gaussian with
respect to a normalised vector $\psi _{0}\in \mathrm{D}$, $\Vert \psi
_{0}\Vert ^{2}=(\psi _{0}|\psi _{0})=1$ if 
\begin{equation}
\big(\psi _{0}\mid e^{R(i\xi )}\psi _{0}\big)=e^{\vartheta _{0}(i\xi )-\xi
^{2}/2}\equiv \theta _{0}(\xi ),\qquad \xi \in \func{Re}\,\Xi .
\label{eq:diff3.2}
\end{equation}%
Here $\vartheta _{0}(\eta )=\langle \eta ,\vartheta _{0}\rangle $ is the
linear continuous $\sharp -$functional\break $\vartheta _{0}(\eta )^{\ast
}=\vartheta _{0}(\eta ^{\sharp })$ of the mathematical expectation $(\psi
_{0}|R(\eta )\,\psi _{0})=\vartheta _{0}(\eta )$, that is defined by some $%
\vartheta _{0}\in \func{Re}\,\Xi $ by means of the complexified bilinear
form 
\begin{equation*}
\langle \xi +i\eta ,\vartheta _{0}\rangle =\langle \xi ,\vartheta
_{0}\rangle +i\langle \eta ,\vartheta _{0}\rangle
\end{equation*}%
on $\Xi $. The product $\langle \xi ,\eta \rangle $ corresponds to the
symmetric covariance 
\begin{equation*}
\func{Re}\,\big(R(\xi )\,\psi _{0}\mid R(\eta )\,\psi _{0}\big)-\vartheta
_{0}(\xi )\,\vartheta _{0}(\eta )=\langle \xi ,\eta \rangle ,\qquad \forall
\xi ,\eta \in \func{Re}\,\Xi .
\end{equation*}%
The exponents $e^{R(i\xi )}=X(\xi )=e^{iR(\xi )}$ are defined as unitary
operators, which form the Weyl family $\{X(\xi )|\xi \in \func{Re}\,\Xi \}$, 
\begin{equation*}
X(\xi )\,X(\eta )=e^{is(\xi ,\eta )}X(\xi +\eta ),\qquad \forall \,\xi ,\eta
\in \func{Re}\,\Xi ,
\end{equation*}%
with the self-adjoint generators 
\begin{equation*}
R(\xi )\,\psi =-i{\frac{\mathrm{d}}{\mathrm{d}\lambda }}\,X(\lambda \xi
)\,\psi |_{\lambda =0},\qquad \xi \in \func{Re}\,\Xi .
\end{equation*}

Such a representation can be realised in the Fock space $\mathrm{H}=\mathrm{F%
}$ over the completion $\mathrm{K\ }$ of the (quotient) space $\Xi $ with
respect to the (semi) positive definite scalar product 
\begin{equation*}
(\xi \mid \eta )=\langle \eta ,\xi ^{\sharp }\rangle +{\frac i2}\,s(\eta
,\xi ^{\sharp }),\qquad \forall \,\xi ,\eta \in \Xi .
\end{equation*}

Indeed, the space $\mathrm{F\ }$can be defined as the comletion of the
(quotient) span $\mathrm{D\ }$of the exponential vectors $\{\eta ^{\otimes
}\mid \eta \in \Xi \}$ with respect to the scalar product%
\begin{equation*}
(\xi ^{\otimes }\mid \eta ^{\otimes })=\sum_{n=0}^{\infty }\frac{1}{n!}(\xi
\mid \eta )^{n}=\exp (\xi \mid \eta ).
\end{equation*}%
Let $A^{\ast }:\Xi \rightarrow \mathcal{B}(\mathrm{D})$ be a linear map,
defining the creation operators in $\mathrm{D}$ by the adjoints $A^{\ast
}(\eta )=A(\eta ^{\sharp })^{\ast }$ to the annihilation operators 
\begin{equation*}
A(\xi ):\eta ^{\otimes }\mapsto (\xi \mid \eta )\eta ^{\otimes },\qquad
\forall \,\xi \in \func{Re}\Xi .
\end{equation*}%
From the canonical commutation relations 
\begin{equation*}
\big[A(\eta ^{\sharp }),A^{\ast }(\eta )\big]=(\eta \mid \eta )\,I\geq
0,\qquad \forall \,\eta \in \Xi ,
\end{equation*}%
one can obtain the relations (\ref{eq:diff3.1}) for the linear combinations 
\begin{equation*}
R(\eta )=\vartheta _{0}(\eta )\,I+2\Re \,A(\eta ),\qquad 2\Re \,A(\eta
)=A^{\ast }(\eta )+A(\eta ).
\end{equation*}%
This defines the Gaussian representation $\xi \mapsto $$R(\xi )$ with
respect to the vacuum vector $\psi _{0}=0^{\otimes }$ in $\mathrm{F}$, so
that $A(\eta )\,\psi _{0}=0,\ \forall \,\eta $. The Weyl operators $X(\xi )$
are defined in $\mathrm{F}$ as 
\begin{equation}
X(\xi )=\theta _{0}(\xi )\,e^{A^{\ast }(i\xi )}e^{A(i\xi )},\qquad \forall
\,\xi \in \func{Re}\,\Xi .  \label{eq:diff3.3}
\end{equation}%
We obtain the representation (\ref{eq:diff3.2}): $\theta _{0}(\xi )=(\psi
_{0}|X(\xi )\,\psi _{0})$ if we take into account the fact that $e^{A(\eta
)}\psi _{0}=\psi _{0}$.

Let us denote by $\mathbf{j}:\eta \mapsto \mathbf{j}\eta (=\eta )$ a
canonical bounded map from the Hilbert space $\Xi $ with respect to the norm 
\begin{equation*}
|\eta |=\langle \eta ^{\sharp },\eta \rangle ^{1/2}=\sqrt{(\func{Re}\,\eta
)^{2}+(\func{Im}\,\eta )^{2}},\qquad \func{Im}\,\eta ^{\sharp }=\func{Re}%
\,i\eta ,
\end{equation*}%
into the pre-Hilbert space $\Xi $ with respect to the (semi) norm $\Vert
\eta \Vert =(\eta \mid \eta )^{1/2}$, 
\begin{eqnarray*}
\Vert \eta \Vert ^{2} &=&(\eta \mid \eta )=|\eta |^{2}+{\frac{i}{2}}%
\,s\,(\eta ,\eta ^{\sharp })=|\eta |^{2}+s\,(\func{Re}\eta ,\func{Im}\eta )
\\
&\leq &|\eta |^{2}+\big|s\,(\func{Re}\eta ,\func{Im}\eta )\big|\leq |\eta
|^{2}+2\,|\func{Re}\eta |\,|\func{Im}\eta |\leq 2|\eta |^{2}.
\end{eqnarray*}%
Then we can write $(\xi |\eta )=\langle \xi ^{\sharp },\mathbf{g}\eta
\rangle $, where $\mathbf{g}=\mathbf{1}-{\frac{i}{2}}\mathbf{s}$, so that $%
\mathbf{g}\eta $ is the complex bounded functional $\vartheta (\xi )=(\xi
^{\sharp }|\eta )=\langle \xi ,\mathbf{g}\eta \rangle $ on $\func{Re}\Xi $
which together with $\vartheta ^{\sharp }(\xi )=\langle \xi ^{\sharp },%
\mathbf{g}\eta \rangle ^{\ast }=(\eta |\xi )$ defines the Hermitian
functional 
\begin{equation*}
2\func{Re}\vartheta =\vartheta +\vartheta ^{\sharp }=2\func{Re}\eta +\mathbf{%
s}\func{Im}\eta ,
\end{equation*}%
where $\mathbf{s}:\func{Re}\Xi \rightarrow \func{Re}\Xi $ is a
skew-symmetric operator $\langle \xi ,\mathbf{s}\eta \rangle =s(\eta ,\xi )$%
, $|\mathbf{s}\eta |\leq 2|\eta |$.

Let us consider a quantum diffusion for the operators $R(t,\eta )=\pi
(t,R(\eta ))$ with continuous indirect sequential observation of the
operators 
\begin{equation*}
G_{t}=L_{t}+L_{t}^{\ast }=R(\zeta _{t}+\zeta _{t}^{\sharp }).
\end{equation*}%
Here $L_{t}=R(\zeta _{t})$, $L_{t}^{\ast }=R(\zeta _{t}^{\sharp })$ are
defined by the weak locally square-integrable families $\{\zeta _{t}\}$, $%
\{\zeta _{t}^{\sharp }\}$ of the elements $\zeta _{t},\ \zeta _{t}^{\sharp
}\in \Xi $, $t\in \mathbf{R}_{+}$ so that%
\begin{equation*}
\int_{0}^{t}\varepsilon _{r}(\vartheta ^{\sharp },\vartheta )\mathrm{d}%
r<\infty ,\forall \,\vartheta \in \Xi ,t\in \mathbf{R}_{+},
\end{equation*}%
where 
\begin{equation}
\varepsilon _{t}(\vartheta ^{\sharp },\vartheta )=\vartheta ^{\sharp }(\zeta
_{t})\,\vartheta (\zeta _{t}^{\sharp })=\big|\langle \vartheta ^{\sharp
},\zeta _{t}\rangle \big|^{2}.  \label{eq:diff3.4}
\end{equation}%
Taking into account the fact that any derivation of the Weyl algebra $%
\mathcal{A}$ is internal, we consider only the structural maps (\ref%
{eq:diff1.10}) , given by the operators $L_{t},L_{t}^{\ast }$ and by a
Hamiltonian $H_{t}$ in the initial Fock space $\mathrm{H}=\mathrm{F}$ .
Moreover, we shall assume that the Hamiltonian is obtained by the normal
ordering $H_{t}=:h_{t}(R):$ of a quadratic form $h_{t}(R)$ of the operators $%
A+A^{\ast }$ such that 
\begin{equation*}
\big(\psi _{\eta }\mid H_{t}\psi _{\eta }\big)=\upsilon _{t}(\vartheta )+{%
\frac{1}{2}}\,\omega _{t}(\vartheta ,\vartheta ),\qquad \vartheta =\vartheta
_{0}+2\func{Re}(\mathbf{g}\eta ).
\end{equation*}%
In this equation $\psi _{\eta }=\exp \{-{\frac{1}{2}}(\eta |\eta )+A^{\ast
}(\eta )\}\,\psi _{0}$ are the normalised exponential vectors 
\begin{equation*}
\psi _{\eta }=e^{-\Vert \eta \Vert ^{2}/2}\eta ^{\otimes },\qquad \Vert \psi
_{\eta }\Vert ^{2}=(\psi _{\eta }\mid \psi _{\eta })=1,\quad \forall \,\eta
\in \Xi ,
\end{equation*}%
generating $\mathrm{H}=\mathrm{F}$ as the completion of the linear envelope $%
\{\psi _{\eta }\}$. $\{\upsilon _{t}|t\in \mathbf{R}_{+}\}$ is a locally
integrable family of the linear $\sharp -$forms $\upsilon _{t}:\Xi
\rightarrow \mathbf{C}$, and $\{\omega _{t}|t\in \mathbf{R}_{+}\}$ is a
locally integrable family of real symmetric forms 
\begin{equation*}
\omega _{t}(\vartheta ^{\sharp },\vartheta )=\omega _{t}(\func{Re}\vartheta ,%
\func{Re}\vartheta )+\omega _{t}(\func{Im}\vartheta ,\func{Im}\vartheta
)=\omega _{t}(\vartheta ,\vartheta ^{\sharp }).
\end{equation*}%
We shall consider the forms $\upsilon _{t},\omega _{t}$ to be continuous, so
that 
\begin{equation*}
\upsilon _{t}(\vartheta )=\langle \upsilon _{t},\vartheta \rangle ,\qquad
\omega _{t}(\vartheta ^{\prime },\vartheta )=\vartheta ^{\prime }(%
\boldsymbol{\omega }_{t}\vartheta )=\langle \vartheta ^{\prime },\boldsymbol{%
\omega }_{t}\vartheta \rangle ,\quad \forall \,\vartheta ,\vartheta ^{\prime
}\in \Xi ,
\end{equation*}%
where $\upsilon _{t}\in \func{Re}\Xi $ and $\boldsymbol{\omega }_{t}$ is a
symmetric operator which is bounded with respect to the norm in $\Xi $.

\noindent \textbf{Proposition 2} Under the assumptions made above about the
linearity of $L_{t}$, $L_{t}^{\ast }$ in $\{R(\eta )\}$ and quadraticity of $%
H_{t}$ the equation (\ref{eq:diff1.3}) for $X(t)=R(t,\eta )$, $\eta \in \Xi $
with $D=[X,L],\ D^{\ast }=[L^{\ast },X]$ and $C$ defined by the generator (%
\ref{eq:diff1.10}), is linear and autonomous with respect to the family $%
\{R(\eta )|\eta \in \Xi \},$ 
\begin{equation}
\mathrm{d}R(t,\eta )+R(t,i\boldsymbol{\kappa }_{t}\mathbf{s}\eta )\,\mathrm{d%
}t=\mathrm{d}\widehat{v}_{t}(\eta )+I\upsilon _{t}(\mathbf{s}\eta )\,\mathrm{%
d}t,  \label{eq:diff3.5}
\end{equation}%
where $\boldsymbol{\kappa }_{t}$ is a complex bounded operator in $\Xi $ and 
\begin{equation*}
\mathrm{d}\widehat{v}_{t}(\eta )=s(\eta ,\func{Im}\zeta _{t})\,\mathrm{d}%
\widehat{w}_{t}+s\,(\eta ,2\func{Re}\zeta _{t})\,\mathrm{d}\widehat{u}_{t}={%
\frac{1}{2}}\,\mathrm{d}\widehat{w}_{t}\big(s(\zeta ,i\eta )\big),
\end{equation*}%
where $\widehat{u}_{t}=\Im A_{t}^{\ast }$.

Indeed, for the operators $H_t$ with the quadratic Wick symbols $%
h_t(\vartheta )=\upsilon _t(\vartheta )+{\frac 12}\,\omega _t(\vartheta
,\vartheta )$ the following commutation relations hold: 
\begin{equation*}
i\big[H_t,R(\eta )\big]=\upsilon _t(\mathbf{s}\eta )\,I+R(\omega _t\mathbf{s}%
\eta ),\qquad \forall \,\eta \in \Xi .
\end{equation*}

In addition, the equations (\ref{eq:diff3.1}) also give 
\begin{equation*}
i\big[R(\eta ),L_{t}\big]=s(\eta ,\zeta _{t})\,I,\qquad i\big[L_{t}^{\ast
},R(\eta )\big]=s(\zeta _{t}^{\sharp },\eta )\,I
\end{equation*}%
for $L_{t}=R(\zeta _{t}),\ L_{t}^{\ast }=R(\zeta _{t}^{\sharp })$. By
substituting this into (\ref{eq:diff1.10}) we obtain $C_{t}=R(i\boldsymbol{%
\kappa }_{t}\mathbf{s}\eta )-\upsilon _{t}(\mathbf{s}\eta )\,I$, where 
\begin{equation*}
\boldsymbol{\kappa }_{t}={\frac{1}{2}}\,\boldsymbol{\gamma }_{t}+i%
\boldsymbol{\omega }_{t},\qquad \boldsymbol{\gamma }_{t}=\boldsymbol{%
\varepsilon }_{t}-\boldsymbol{\varepsilon }_{t}^{\sharp }.
\end{equation*}%
Here$\boldsymbol{\varepsilon }_{t}$~is an operator in $\Xi $, that defines
the positive form (\ref{eq:diff2.4}) 
\begin{equation*}
\varepsilon _{t}(\vartheta ^{\sharp },\vartheta )=\langle \vartheta ^{\sharp
},\mbox{\boldmath$\varepsilon$}_{t}\vartheta \rangle ,\qquad \langle
\vartheta ^{\sharp },\mbox{\boldmath$\varepsilon$}_{t}^{\sharp }\vartheta
\rangle ,=\varepsilon _{t}(\vartheta ,\vartheta ^{\sharp }),\qquad \forall
\,\vartheta \in \Xi .
\end{equation*}%
In the right-hand side of the equation (\ref{eq:diff1.3}) we obtain the
quantum noise 
\begin{eqnarray}
D_{t}^{\ast }\mathrm{d}A_{t}+D_{t}\mathrm{d}A_{t}^{\ast } &=&\big[R(\zeta
_{t}^{\sharp }),R(\eta )\big]\,\mathrm{d}A_{t}+\big[R(\eta ),R(\zeta _{t})%
\big]\,\mathrm{d}A_{t}^{\ast }  \notag \\
&=&i\,\big(s(\eta ,\zeta _{t}^{\ast })\,\mathrm{d}A_{t}-s(\eta ,\zeta _{t})\,%
\mathrm{d}A_{t}^{\ast }\big)=\mathrm{d}\widehat{v}_{t}(\eta ).
\label{eq:diff3.6}
\end{eqnarray}

The following theorem establishes the existence and uniqueness of the
solution of equation (\ref{eq:diff3.5}) together with the integral of the
locally square-integrable real function $g(t)$ with respect to the
stochastic differentials 
\begin{equation}  \label{eq:diff3.6'}
\mathrm{d}Y(t)=R(t,\zeta _t+\zeta _t^{\sharp })\,\mathrm{d}t+\mathrm{d}W_t,
\end{equation}
where $W_t=A_t+A_t^{*}$~is the Fock representation of the Wiener
process~---\ an error of measurement $L_t+L_t^{*}$ in Fock space $\mathcal{F}%
=\Gamma (\mathcal{K})$.

\begin{theorem}
Let the equations (\ref{eq:diff3.5}), (\ref{eq:diff3.6}) for the linear
diffusion be defined by the functions $\upsilon _{t},\omega _{t}$ and $%
\varepsilon _{t}$ , which are locally integrable with respect to the norm in 
$\Xi $, such that 
\begin{equation*}
|\upsilon _{t}|_{t}^{(1)}=\int_{0}^{t}|\upsilon _{r}|\,\mathrm{d}r<\infty
,\qquad |\boldsymbol{\kappa }|_{t}^{(1)}=\int_{0}^{t}|\boldsymbol{\kappa }%
_{r}|\,\mathrm{d}r<\infty ,\quad \forall \,t,
\end{equation*}%
where $|\upsilon _{t}|=\sqrt{\upsilon _{t}^{2}}$, $|\boldsymbol{\kappa }%
_{t}|=\sup \{\kappa _{t}(\vartheta ^{\prime },\vartheta )||\vartheta
^{\prime }|,|\vartheta |\leq 1\}$ , and 
\begin{equation*}
|\zeta +\zeta ^{\sharp }|_{t}^{(2)}=\bigg(\int_{0}^{t}(2\func{Re}\zeta
_{r})^{2}\mathrm{d}r\bigg)^{1/2}<\infty ,\qquad \forall \,t\in \mathbf{R}%
_{+}.
\end{equation*}%
Then they have a unique solution defined in the Hilbert space $\mathrm{F}%
\otimes \mathcal{F}$ as 
\begin{equation}
R(t,\eta )+\int_{0}^{t}g(r)\,\mathrm{d}Y(r)=\int_{0}^{t}s(\upsilon _{r},\eta
_{r})\,\mathrm{d}r+R\big(\eta (t)\big)+\imath _{0}^{t}(f^{\ast },f).,
\label{eq:diff3.7}
\end{equation}%
by the quantum stochastic integral $\imath _{0}^{t}(f^{\ast
},f)=a(f_{t}^{\ast })+a^{\ast }(f_{t})$ with 
\begin{equation*}
a(f_{t}^{\ast })=\int_{0}^{t}f^{\ast }(r,\eta )\,\mathrm{d}A_{r},\qquad
a^{\ast }(f_{t})=\int_{0}^{t}f(r,\eta )\,\mathrm{d}A_{r}^{\ast }.
\end{equation*}%
Here $\eta (t)=\phi _{0}^{(g)}(t,\eta )=\eta _{0},\ \eta _{r}=\phi
_{r}^{(g)}(t,\eta )$, $r\in \lbrack 0,t[$ is the solution of the backward
conjugate equation 
\begin{equation}
-\dot{\eta}_{r}+i\boldsymbol{\kappa }_{r}\mathbf{s}\eta _{r}=g(r)\,(\zeta
_{r}+\zeta _{r}^{\sharp }),  \label{eq:diff3.8}
\end{equation}%
with the boundary condition $\eta _{t}=\eta $. The complex functions $%
f_{t},\ f_{t}^{\ast }$ of $r\in \mathbf{R}_{+}$ and $\eta \in \Xi $ equal
zero at $r>t$ and are defined for $r\leq t$ by the real function $g(r)$ as 
\begin{equation*}
f^{\ast }(r,\eta )=g(r)+is\big(\phi _{r}^{(g)}(t,\eta ),\zeta _{r}^{\sharp }%
\big)=f(r,\eta ^{\sharp })^{\ast }.
\end{equation*}
\end{theorem}

\textsc{Proof} First, we write the weak solution of the equation~(\ref%
{eq:diff3.8}) in the standard Duhamel form 
\begin{equation*}
\eta _{r}=\boldsymbol{\phi }(t)\,\eta +\int_{r}^{t}g(s)\,\boldsymbol{\phi }%
_{r}(s)(\zeta _{s}+\zeta _{s}^{\sharp })\,\mathrm{d}s,
\end{equation*}%
where $\boldsymbol{\phi }_{r}(t)\eta =\boldsymbol{\phi }_{r}^{(0)}(t,\eta )$
is the solution of the equation (\ref{eq:diff3.8}) with zero right-hand side 
$g=0$. The resolving operator $\boldsymbol{\phi }_{r}(t)$ exists as the
chronologically ordered exponential 
\begin{equation*}
\boldsymbol{\phi }_{r}(t)=\sum_{n=0}^{\infty }\mathop{\int\cdots\int}%
\limits_{r\leq t_{1}<\cdots <t_{n}<t}\boldsymbol{\kappa }_{t_{1}}\mathbf{s}%
\cdots \boldsymbol{\kappa }_{t_{n}}\mathbf{s}\mathrm{d}t_{1}\cdots \mathrm{d}%
t_{n}.
\end{equation*}%
This comes from the estimate of the norm 
\begin{eqnarray*}
\big|\boldsymbol{\phi }_{r}(t)\big| &=&\sup \Big\{\big|\boldsymbol{\phi }%
_{r}(t)\,\eta \big|:\,|\eta |<1\Big\} \\
&\leq &\sum_{n=0}^{\infty }|\mathbf{s}|^{n}\mathop{\int\cdots\int}%
\limits_{0\leq t_{1}<\cdots <t_{n}<t}|\boldsymbol{\kappa }_{t_{1}}|\cdots |%
\boldsymbol{\kappa }_{t_{n}}|\mathrm{d}t_{1}\cdots \mathrm{d}t_{n} \\
&\leq &\exp \bigg\{\int_{0}^{t}|2\boldsymbol{\kappa }_{r}|\,\mathrm{d}r%
\bigg\},
\end{eqnarray*}%
which is finite as $|\mathbf{s}|\leq 2$. Hence, the linear form $\langle
\eta _{r},\vartheta \rangle $ is uniquely defined on $\Xi \ni \vartheta $
for every $r\in \lbrack 0,t)$ as a bounded functional with the estimate {%
\begin{eqnarray*}
\big|\langle \eta _{r},\vartheta \rangle \big| &\leq &\big|\langle 
\boldsymbol{\phi }_{r}(t),\vartheta \rangle \big|+\int_{r}^{t}\big|g(s)\big|%
\langle \boldsymbol{\phi }_{r}(s)\,2\func{Re}\zeta _{s},\vartheta \rangle %
\big|\,\mathrm{d}s \\
&\leq &|\eta |\,\big|\boldsymbol{\phi }_{r}^{\intercal }(t)\,\vartheta \big|%
+|g|_{t}^{(2)}|\zeta +\zeta ^{\sharp }|_{t}^{(2)}\bigg(\int_{r}^{t}\big|%
\boldsymbol{\phi }_{r}^{\intercal }(s)\,\vartheta \big|^{2}\mathrm{d}s\bigg)%
^{1/2} \\
&\leq &\Big(|\eta |+|g|_{t}^{(2)}|\zeta +\zeta ^{\sharp }|_{t}^{(2)}\sqrt{t-r%
}\Big)\,|\vartheta |\,\exp \big\{2\,|\boldsymbol{\kappa }|_{t}^{(1)}\big\}.
\end{eqnarray*}%
}

Now we can integrate the left-hand side of the equation (\ref{eq:diff3.7})
by parts, taking into account (\ref{eq:diff3.6}) and (\ref{eq:diff3.8}): 
\begin{eqnarray*}
R(t,\eta ) &+&\int_{0}^{t}g(r)\,\big(R(r,2\func{Re}\zeta _{r})\,\mathrm{d}%
r+2\Re \mathrm{d}A_{r}\big) \\
&=&R(t,\eta )+\int_{0}^{t}\big(2\Re \big\{g(r)\,\mathrm{d}A_{r}\big\}-R\,(r,%
\dot{\eta}_{r}-i\boldsymbol{\kappa }_{r}\mathbf{s}\eta _{r})\,\mathrm{d}r%
\big) \\
&=&R(0,\eta _{0})+\int_{0}^{t}\big(2\Re \big\{g(r)\,\mathrm{d}A_{r}\big\}+%
\mathrm{d}R\,(r,\eta _{r})+R(r,i\boldsymbol{\kappa }_{r}\mathbf{s}\eta
_{r})\,\mathrm{d}r\big) \\
&=&R\big(\phi _{0}^{(g)}(t,\eta )\big)+\int_{0}^{t}\big(2\Re \big\{%
g(r)+is\,(\eta _{r},\zeta _{r}^{\sharp })\,\mathrm{d}A_{r}\big\}+\upsilon
_{r}(\mathbf{s}\eta _{r})\,\mathrm{d}r\big).
\end{eqnarray*}%
Here $\mathrm{d}R(r,\eta _{r})$ is the quantum stochastic differential $%
\mathrm{d}R(r,\eta )|_{\eta =\eta _{r}}$, satisfying the equation (\ref%
{eq:diff3.5}) for $t=r$. This proves Theorem~2.

\noindent \textbf{Remark 2} The solution $R(t,\eta )$ of the equation (\ref%
{eq:diff3.5}) given by the integral (\ref{eq:diff3.7}) for $g=0$ preserves
the commutation relations (\ref{eq:diff3.1}) and satisfies the nondemolition
principle 
\begin{equation*}
\big[R(t,\eta ),Y_g(t)\big]=0,\qquad \forall \,\eta \in \Xi ,\quad g\in L^2(%
\mathbf{R}_{+}),
\end{equation*}
with respect to the commutative (self-nondemolition) processes 
\begin{equation*}
Y_g(t)=\int_0^tg(r)\,\mathrm{d}Y(r),\qquad g\in L^2(\mathbf{R}_{+}).
\end{equation*}

Indeed, by using the quantum Ito formula we can obtain 
\begin{eqnarray*}
\big[\mathrm{d}R(t,\eta ),\mathrm{d}R(t,\eta ^{\sharp })\big] &=&\big[%
\mathrm{d}\widehat{v}_{t}(\eta ),\mathrm{d}\widehat{v}_{t}(\eta ^{\sharp })%
\big]=\gamma _{t}(\mathbf{s}\eta ,\mathbf{s}\eta ^{\sharp })\,\mathrm{d}t, \\
\big[\mathrm{d}R(t,\eta ),\mathrm{d}Y_{g}(t)\big] &=&\big[\mathrm{d}\widehat{%
v}_{t}(\eta ),\mathrm{d}\widehat{w}_{t}g(t)\big]=is(\eta ,\zeta +\zeta
^{\sharp })\,g(t)\,\mathrm{d}t.
\end{eqnarray*}%
Hence, if $[R(t,\eta ),R(t,\eta ^{\sharp })]=(1/i)s\,(\eta ,\eta ^{\sharp
})\,I$, then 
\begin{eqnarray*}
\big[\mathrm{d}R(t,\eta ),R(t,\eta ^{\sharp })\big] &=&\kappa
_{t}^{\intercal }(\mathbf{s}\eta ,\mathbf{s}\eta ^{\sharp })\mathrm{d}t, \\
\big[R(t,\eta ),\mathrm{d}R(t,\eta ^{\sharp })\big] &=&-\kappa _{t}(\mathbf{s%
}\eta ,\mathbf{s}\eta ^{\sharp })\mathrm{d}t
\end{eqnarray*}%
and 
\begin{eqnarray*}
\mathrm{d}\big[R(t,\eta ),R(t,\eta ^{\sharp })\big] &=&\kappa
_{t}^{\intercal }-\kappa _{t}+\gamma _{t})\,(\mathbf{s}\eta ,\mathbf{s}\eta
^{\sharp })\,\mathrm{d}t=0, \\
\mathrm{d}\big[R(t,\eta ),Y_{g}(t)\big] &=&\big[\mathrm{d}R(t,\eta ),Y_{g}(t)%
\big]+\big[R(t,\eta ),\mathrm{d}Y_{g}(t)\big]+\big[\mathrm{d}R(t,\eta
),Y_{g}(t)\big] \\
&=&\Big\{\big[R(t,\eta ),R(t,\zeta _{t}+\zeta _{t}^{\sharp })\big]+is(\eta
,\zeta _{t}+\zeta _{t}^{\sharp })\,I\Big\}\,g(t)\,\mathrm{d}t=0,
\end{eqnarray*}%
if $[R(t,\eta ),\,Y_{g}(t)]=0$ and, consequently, $[\mathrm{d}R(t,\eta
),\,Y_{g}(t)]=0$.

\noindent \textbf{Example.\/} Let us consider the simplest nontrivial case
of the space $\Xi =\mathbf{C}^{2}$ of column vectors $\eta =\left[ 
\begin{array}{c}
\eta _{p} \\ 
\eta _{q}%
\end{array}%
\right] $, $\eta _{p},\eta _{q}\in \mathbf{C}$ with the involution $\eta
\mapsto \eta ^{\sharp }$ (a complex conjugation $\eta ^{\sharp }=\left[ 
\begin{array}{c}
\eta _{p}^{\ast } \\ 
\eta _{q}^{\ast }%
\end{array}%
\right] $) and nondegenerate symplectic form 
\begin{equation*}
s(\eta ,\eta ^{\sharp })=2(\eta _{p}\eta _{q}^{\ast }-\eta _{q}\eta
_{p}^{\ast }).
\end{equation*}%
This corresponds to the canonical commutation relations $[P,Q]=(2/i)I$ for
the operators momentum $P$ and coordinate $Q$ for the one-dimensional
quantum particle. They are defined in the Fock space over $\mathrm{K}=%
\mathbf{C}$ as (\ref{eq:diff3.1}) for $R(\eta )=\eta _{p}P+\eta _{q}Q$ due
to the degeneracy of the semi positive definite scalar product 
\begin{equation*}
(\xi \mid \eta )=\eta _{p}\xi _{p}^{\ast }+\eta _{q}\xi _{q}^{\ast }+{\frac{i%
}{2}}\,s(\eta ,\xi ^{\sharp })=(\xi _{p}^{\ast }+i\xi _{q}^{\ast })\,(\eta
_{p}-i\eta _{q}).
\end{equation*}%
This Fock representation in \textrm{F}$=\Gamma (\mathbf{C})$ is associated
with the Gaussian state (\ref{eq:diff3.2}) (which corresponds to the
standard scalar product $\langle \xi ^{\sharp },\eta \rangle =\xi _{p}^{\ast
}\xi _{p}+\xi _{q}^{\ast }\xi _{q}$ ) and is equivalent to the Shr\"{o}%
dinger representation in $\mathrm{H}=L^{2}(\mathbf{R})$. In this
representation $Q_{0}=x$ and $P_{0}={\frac{2}{i}}{\frac{\mathrm{d}}{\mathrm{d%
}x}}$, and 
\begin{equation*}
\psi _{0}(x)={\frac{1}{(2\pi )^{1/4}}}\,\exp \Big\{-{\frac{1}{4}}(x-q)^{2}+{%
\frac{i}{2}}\,qx\Big\},
\end{equation*}%
where $q=(\psi _{0}|Q\psi _{0})=\vartheta _{q}$ and $p=(\psi _{0}|P\psi
_{0})=\vartheta _{q}$.

Suppose that this particle exhibits free quantum Brownian motion, i.e. $%
H=(1/(2m))P^{2}$ is its Hamiltonian, where $m>0$ is the mass of the
particle, so that 
\begin{eqnarray*}
P(t) &=&P\otimes \hat{1}+\widehat{v}_{t},\qquad Q(t)=Q\otimes \hat{1}+{\frac{%
1}{m}}\,\int_{0}^{t}P(r)\,\mathrm{d}r, \\
\widehat{v}_{t} &=&i\sqrt{\lambda }(A_{t}^{\ast }-A_{t})=I\otimes 2\sqrt{%
\lambda }\Im A_{t}^{\ast }
\end{eqnarray*}%
is a realisation in the Fock space $\mathrm{F}\otimes \mathcal{F}$, where $%
\mathcal{F}=\Gamma (L^{2}(\mathbf{R}^{+}))$, of the Wiener process with
intensity $2\lambda $ with respect to the vacuum-vector $\delta _{\emptyset
}\in \mathcal{F}$. This corresponds to the quantum stochastic equation (\ref%
{eq:diff3.5}) with the parameters $\upsilon _{t}=0$, 
\begin{equation*}
\boldsymbol{\omega }_{t}=\left( 
\begin{array}{cc}
1/2m & 0 \\ 
0 & 0%
\end{array}%
\right) ,\qquad \mathbf{s}_{t}=\left( 
\begin{array}{cc}
0 & -2 \\ 
2 & 0%
\end{array}%
\right) ,\qquad \zeta _{t}={\frac{\sqrt{\lambda }}{2}}\,\left( 
\begin{array}{c}
0 \\ 
1%
\end{array}%
\right) =\zeta _{t}^{\sharp }.
\end{equation*}%
In this case the equation (\ref{eq:diff3.6}) describes an indirect
observation of the coordinate of the particle, realised by measuring the
increment of the commutative process 
\begin{equation*}
Y(t)=\int_{0}^{t}\sqrt{\lambda }Q(r)\,\mathrm{d}r+W_{t},\qquad Y(0)=0,
\end{equation*}%
with precision $\lambda >0$, where $W_{t}=A_{t}+A_{t}^{\ast }$ is the
standard stochastic error of the measurement represented by another
realisation in \textrm{F}$\otimes \mathcal{F}$ of the standard Wiener
process with respect to $\delta _{\emptyset }$. Note, that by the
noncommutativity 
\begin{equation*}
\big[\widehat{v}(s),\widehat{w}(r)\big]=2i\sqrt{\lambda }\,\min
\{s,r\}\,I\otimes \hat{1}
\end{equation*}%
it is impossible to represent these quantum processes together as Wiener
classical processes on the same probability space, although each of them
separately allows such a representation. In addition, the nondemolition
condition of the pair $(P,Q)$ with respect to the observation $Y$ can be
verified directly by calculating the commutators 
\begin{equation*}
\big[P(t),Y(s)\big]=0,\qquad \big[Q(t),Y(s)\big]=0,\qquad \forall \,s\leq t.
\end{equation*}%
(One can check, that for $s>t$ the above do not commute)

\section{Markovian filtering of Gaussian quantum process}

\setcounter{equation}{0}

The solution (\ref{eq:diff3.7}), obtained for the linear quantum diffusion
equation (\ref{eq:diff3.5}) with continuous observation (\ref{eq:diff3.6}),
enables us to find the Weyl operators (\ref{eq:diff3.3}) in the Heisenberg
picture $X(t,\xi )=\iota (t,X(\xi ))$. To this end let us represent the
product 
\begin{equation*}
X_{g}(t,\xi )=\exp \bigg\{\int_{0}^{t}\Big(g(r)\,\mathrm{d}Y(r)-{\frac{1}{2}}%
\,g(r)^{2}\mathrm{d}r\Big)\bigg\}\,X(t,\xi )
\end{equation*}%
as the exponent of the operator (\ref{eq:diff3.7}): 
\begin{eqnarray*}
X_{g}(t,\xi ) &=&\exp \bigg\{R(t,i\xi )+\int_{0}^{t}\bigg(g(r)\,\mathrm{d}%
Y(r)-{\frac{1}{2}}\,g(r)^{2}\mathrm{d}r\bigg)\bigg\} \\
&=&\exp \bigg\{\int_{0}^{t}\bigg(s\big(\upsilon _{t},\phi _{r}^{(g)}(t)\big)-%
{\frac{1}{2}}\,g(r)^{2}\bigg)\,\mathrm{d}r \\
&+&R\big(\phi _{0}^{(g)}(t)\big)+\imath _{0}^{t}(f^{\ast },f)\bigg\}\,(i\xi
).
\end{eqnarray*}%
Taking into account that 
\begin{equation*}
\exp \big\{\imath _{0}^{t}(f^{\ast },f)\big\}=e^{|f_{t}|^{2}/2}e^{a^{\ast
}(f_{t})}e^{a(f_{t}^{\ast })},
\end{equation*}%
we can obtain 
\begin{equation*}
X_{g}(t,\xi )=e^{I_{g}^{t}(\xi )}e^{a^{\ast }(f_{t}(i\xi ))}X\Big({\frac{1}{i%
}}\,\phi _{0}^{(g)}(t,i\xi )\Big)\,e^{a(f_{t}(i\xi ))},
\end{equation*}%
where $f_{t}(r,i\xi )=g(r)-is(\eta _{r},\zeta _{t})$, $f_{t}^{\ast }(r,i\xi
)=g(r)+is(\eta _{r},\zeta _{t}^{\sharp })$ for $\eta _{t}=i\xi $. The
integral over the trajectories of the equation (\ref{eq:diff3.8}) 
\begin{equation*}
I_{g}^{t}(\xi )=\int_{0}^{t}s\big(\upsilon _{r},\phi _{r}^{(g)}(t,i\xi )\big)%
\,\mathrm{d}r+{\ \frac{|f_{t}|^{2}(i\xi )-g_{t}^{2}}{2}},
\end{equation*}%
where 
\begin{equation*}
g_{t}^{2}=\int_{0}^{t}g(r)^{2}\mathrm{d}r,\qquad |f_{t}|^{2}(i\xi
)=\int_{0}^{t}f^{\ast }(r,i\xi )\,f(r,i\xi )\,\mathrm{d}r,
\end{equation*}%
can be written using (\ref{eq:diff3.4}) in the following form 
\begin{equation}
I_{g}^{t}(\xi )=\int_{0}^{t}\Big\{s\big(\upsilon _{r}+\func{Im}\zeta
_{r}^{\sharp }g(r),\eta _{r}\big)+{\frac{1}{2}}\,\varepsilon _{r}(\mathbf{s}%
\eta _{r},\mathbf{s}\eta _{r})\Big\}\,\mathrm{d}r.  \label{eq:diff4.1}
\end{equation}%
Given that $e^{a(f^{\ast })}\delta _{\emptyset }=0$ we deduce the following
result.

\noindent \textbf{Proposition 3\/} The mathematical expectation $(\psi
\otimes \delta _\emptyset |X_g(t,\xi )\,\psi \otimes \delta _\emptyset )$
with respect to the state vector $h=\psi \otimes \delta _\emptyset $ for
arbitrary $\psi \in \mathrm{F}$ can be written in the form $\langle \psi
|\chi _g^t(\xi )\,\psi \rangle $, where the expectation $\chi _g^t(\xi )$ of
the operator $X_g(t,\xi )$ with respect to the vacuum vector $\delta
_\emptyset \in \mathcal{F}$ has the form 
\begin{equation}  \label{eq:diff4.2}
\chi _g^t(\xi )=\exp \Big\{I_g^t(\xi )+R\big(\phi _0^{(g)}(t,i\xi )\big)%
\Big\}.
\end{equation}

The function $t\mapsto \chi _{g}^{t}(\xi )$ satisfies the operator equation 
\begin{eqnarray}
i\,{\frac{\mathrm{d}}{\mathrm{d}t}}\,\chi _{g}^{t}(\xi ) &+&\Big\{s(\upsilon
_{t},\xi )+{\frac{i}{2}}\,\varepsilon _{t}(\mathbf{s}\xi ,\mathbf{s}\xi )%
\Big\}\,\chi _{g}^{t}(\xi )  \notag \\
&=&\big\{\langle \boldsymbol{\kappa }_{t}\mathbf{s}\xi ,\partial \rangle +2%
\func{Im}\langle \zeta _{t}g(t),i\partial +{\frac{1}{2}}\,\mathbf{s}\xi
\rangle \big\}\,\chi _{g}^{t}(\xi ),  \label{eq:diff4.3}
\end{eqnarray}%
in the partial (functional) derivatives of the first order%
\begin{equation*}
\langle \zeta ,\partial \rangle X(\xi )=(i/2)(X(\xi )R(\zeta )+R(\zeta
)X(\xi )).
\end{equation*}
\medskip

Indeed, it is not difficult to find, by means of direct substitution, that
the characteristic operator-function (\ref{eq:diff4.2}) satisfies the
equation (\ref{eq:diff4.3}). However, this can be skipped, as far as this
equation with the commutation relation 
\begin{equation*}
\big[R(\zeta ),X(\xi )\big]=s(\zeta ,\xi )\,X(\xi ),\qquad \forall \,\zeta
,\xi \in \Xi ,
\end{equation*}%
is concerned, it can be written in the form of the equation (\ref{eq:diff2.4}%
), that was derived before for the vacuum expectation $\mu _{g}^{t}(X)$ of
any operator of the form $X_{g}(t)=e_{g}(t)X(t)$: 
\begin{equation*}
{\frac{\mathrm{d}}{\mathrm{d}t}}\,\mu _{g}^{t}(\xi )+\mu _{g}^{t}\Big(\gamma
_{t}\big(X(\xi )\big)\Big)=g(t)\,\mu _{g}^{t}\big(X(\xi )\,R(\zeta
_{t})+R(\zeta _{t}^{\sharp })\,X(\xi )\big),
\end{equation*}%
where 
\begin{equation*}
\gamma _{t}(X)=i\,[X,\upsilon _{t}(R)+{\frac{1}{2}}\,\omega _{t}(R,R)]+{%
\frac{1}{2}}\,(R(\zeta _{t}^{\sharp })[R(\zeta _{t}),X]+[X,R(\zeta
_{t}^{\sharp })]R(\zeta _{t})).
\end{equation*}%
In particular, for $g=0$ this equation defines the Markovian map $\mu
_{0}^{t}:\mathcal{A}\rightarrow \mathcal{A}$ for the quantum diffusion $%
R(t,\eta )$; this map also has the characteristic operator-valued function $%
\chi _{0}^{t}(\xi )=\mu _{0}^{t}(X(\xi ))$ in the form (\ref{eq:diff4.2})
which is defined by the integral $I_{0}^{t}(\xi )$ and $\phi
_{r}^{(0)}(t,i\xi )=i\boldsymbol{\phi }_{r}(t)\xi $.

An a posteriori linear quantum diffusion under the continuous measurement of
the process (\ref{eq:diff3.6}) is described by the a posteriori
characteristic operator-valued function $\widehat{\chi }^{t}(\xi )=\widehat{%
\mu }^{t}(X(\xi ))$. Its Wick symbol is the operator-valued function (\ref%
{eq:diff4.2}) defining the characteristic function 
\begin{equation*}
\theta _{g}^{t}(\xi )=\big(\psi \otimes f^{\otimes }\mid \widehat{\chi }%
^{t}(\xi )\,(\psi \otimes f^{\otimes })\big)\,\exp \big\{-\Vert f\Vert ^{2}%
\big\},\qquad \psi \in \mathrm{H},
\end{equation*}%
of $g(t)=2\func{Re}f(t)$ on the exponential normalised vectors $f^{\otimes
}\exp \{-{\ \frac{1}{2}}\,\Vert f\Vert ^{2}\}$. Consequently, the stochastic
operator-function $\widehat{\chi }^{t}(\xi )$, is obtained (using a
factorial substitution in (\ref{eq:diff4.2}), with the standard Wiener
process $\widehat{w}_{t}$ replacing $\int_{0}^{t}g(r)\mathrm{d}r$ ) along
with the solution $\widehat{\eta }_{r}=\widehat{\phi }_{r}(t,i\xi )$ of the
backward linear stochastic equation 
\begin{equation}
-d_{-}\widehat{\eta }_{r}+i\boldsymbol{\kappa }_{r}\mathbf{s}\widehat{\eta }%
_{r}\mathrm{d}r=(\zeta _{r}+\zeta _{r}^{\sharp })\,\mathrm{d}\widehat{w}%
_{r},\qquad \widehat{\eta }_{t}=i\xi .  \label{eq:diff4.4}
\end{equation}%
Substituting $\eta _{r}=\phi _{r}^{(g)}(t,i\xi )$, defines the solution of
the filtering equation for the case of linear diffusion and $X=X(\xi )$.

In order to find the operator $\widehat{\chi }^{t}(\xi )$ in the form of a
stochastic function $\chi ^{t}(\xi ,\omega )=\omega (\widehat{\chi }^{t}(\xi
))$ (of the trajectories $w_{t}$ of the observable process (\ref{eq:diff3.6}%
)), let us solve the equation (\ref{eq:diff2.6}) for the initial Gaussian
state (\ref{eq:diff3.2}) with an arbitrary $\vartheta (\eta )=\langle \eta
,\vartheta \rangle $ instead of $\vartheta _{0}(\eta )$. In that way we
shall find the Wick symbol 
\begin{equation*}
\widehat{\theta }^{t}(\vartheta ,\xi )=\big(\psi _{\eta }\mid \widehat{\chi }%
^{t}(\xi )\,\psi _{\eta }\big),\qquad \vartheta =\vartheta _{0}+2\func{Re}(%
\mathbf{g}\eta ),
\end{equation*}%
where $\psi _{\eta }=\eta ^{\otimes }\exp \{-{\frac{1}{2}}\,\Vert \eta \Vert
^{2}\}$. This symbol defines $\widehat{\chi }^{t}(\xi )$ with the normal
ordered substitution of $R(\eta )=\vartheta _{0}(\eta )\,I+2\Re A(\eta )$
into $\widehat{\theta }^{t}(\vartheta ,\xi )$ instead of $\vartheta (\eta )$%
. It can be obtained by solving the linear stochastic differential equation
in partial (functional) derivatives of the first order 
\begin{eqnarray}
id\widehat{\theta }^{t}(\xi ) &+&\big\{s(\upsilon _{t},\xi )+{\frac{i}{2}}%
\,\varepsilon _{t}(\mathbf{s}\xi ,\mathbf{s}\xi )\big\}\,\widehat{\theta }%
^{t}(\xi )\,\mathrm{d}t  \notag \\
&=&\big(\langle \boldsymbol{\kappa }_{t}\mathbf{s}\xi ,\partial \rangle \,%
\mathrm{d}t+2\func{Im}\,\langle \zeta _{t}\mathrm{d}\widehat{w}%
_{t},i\partial +{\frac{1}{2}}\,\mathbf{s}\xi \rangle \big)\,\widehat{\theta }%
^{t}(\xi ),  \label{eq:diff4.5}
\end{eqnarray}%
representing the filtering equation in terms of the nonnormalised a
posteriori characteristic functional $\widehat{\theta }^{t}(\xi )=\varphi (%
\widehat{\chi }^{t}(\xi ))$ , for the initial Gaussian characteristic
functional 
\begin{equation*}
\varphi \big(X(\xi )\big)=\exp \Big\{\vartheta (i\xi )-{\frac{1}{2}}\,\xi
^{2}\Big\}=\theta (\xi ).
\end{equation*}

\begin{theorem}
The solution to the filtering equation (\ref{eq:diff4.5}) with the Gaussian
\ $\widehat{\theta }^{0}(\xi )=\theta (\xi )$ defines the stochastici
characteristic functional $\theta ^{t}(\omega )=\omega (\widehat{\theta }%
^{t})$ in the form 
\begin{equation}
\widehat{\theta }^{t}(\xi )=\widehat{\rho }^{t}\exp \bigg\{\widehat{%
\vartheta }_{t}(i\xi )-{\frac{1}{2}}\,p_{t}(\xi ,\xi )\bigg\}.
\label{eq:diff4.6}
\end{equation}%
In this equation $\widehat{\rho }^{t}=\widehat{\theta }^{t}(0)$ is the
Gaussian probability density of the output process (\ref{eq:diff3.6}) with
respect to $\widehat{w}_{t}:$ 
\begin{equation*}
\widehat{\rho }^{t}=\exp \bigg\{\int_{0}^{t}\bigg(\widehat{\vartheta }_{r}%
\big(2\func{Re}\zeta _{r}\big)\,\mathrm{d}\widehat{w}_{r}-{\frac{1}{2}}\,%
\widehat{\vartheta }_{r}(2\func{Re}\zeta _{r})^{2}\mathrm{d}r\bigg)\bigg\},
\end{equation*}%
$\hat{\vartheta}_{t}(\eta )=\langle \eta ,\widehat{\vartheta }_{t}\rangle $
is a linear stochastic functional of the a posteriori mathematical
expectation of the operators $R(t,\eta )$ satisfying the quantum linear
filtering equation 
\begin{equation}
\mathrm{d}\widehat{\vartheta }_{t}(\eta )+\widehat{\vartheta }_{t}(i%
\boldsymbol{\kappa }_{t}\mathbf{s}\eta )\,\mathrm{d}t=2\func{Re}\langle \eta
,\mathbf{k}_{t}\zeta _{t}^{\sharp }\rangle \,\mathrm{d}\widetilde{w}%
_{t}+\upsilon _{t}(\mathbf{s}\eta )\,\mathrm{d}t  \label{eq:diff4.7}
\end{equation}%
with the initial $\widehat{\vartheta }_{0}=\vartheta \in \func{Re}\Xi ,\ 
\mathrm{d}\widetilde{w}_{t}=\mathrm{d}\widehat{w}_{t}-\widehat{\vartheta }%
_{t}\left( 2\func{Re}\zeta _{t}\right) \mathrm{d}t$, $\mathbf{k}_{t}=\mathbf{%
p}_{t}+(i/2)\mathbf{s}$ . Here $p_{t}(\xi ,\eta )=\langle \xi ,\mathbf{p}%
_{t}\eta \rangle $ is a symmetric quadratic form of the a posteriori
covariance $R(t,\eta )$ and $R(t,\xi )$, satisfying the Riccati equation
with the initial $p_{0}(\xi ,\eta )=\langle \xi ,\eta \rangle :$ 
\begin{equation}
{\frac{\mathrm{d}}{\mathrm{d}t}}\,p_{t}(\eta ,\eta )-2p_{t}(\eta ,i%
\boldsymbol{\kappa }_{t}\eta )=\varepsilon _{t}(\mathbf{s}\eta ,\mathbf{s}%
\eta )-\big|2\func{Re}\langle \eta ,\mathbf{k}_{t}\zeta ^{\sharp }\rangle %
\big|^{2}.  \label{eq:diff4.8}
\end{equation}
\end{theorem}

\textsc{Proof} Let us find from (\ref{eq:diff4.5}) the stochastic equation
for $\widehat{\sigma }^{t}(i\xi ,\omega )=\ln \widehat{\theta }^{t}(\xi
,\omega )$ using the logarithmic Ito formula 
\begin{equation*}
\mathrm{d}\widehat{\sigma }^{t}(i\xi )={\frac{1}{\widehat{\theta }^{t}(\xi )}%
}\,\mathrm{d}\widehat{\theta }^{t}(\xi )-{\ \frac{1}{2}}\,\big(\mathrm{d}%
\widehat{\sigma }^{t}(i\xi )\big)^{2},
\end{equation*}%
where 
\begin{eqnarray*}
\big(\mathrm{d}\widehat{\sigma }^{t}(\eta )\big)^{2} &=&\bigg(\mathrm{d}%
\widehat{\theta }^{t}\Big({\frac{1}{i}}\,\eta \Big)\Big/\widehat{\theta }^{t}%
\Big({\frac{1}{i}}\,\eta \Big)\bigg)^{2} \\
&=&\Big(\big\langle2\func{Re}\zeta _{t},\partial \widehat{\sigma }^{t}(\eta )%
\big\rangle+\langle \func{Im}\zeta ^{\sharp },\mathbf{s}\eta \rangle \Big)%
^{2}\mathrm{d}t \\
&=&\Big\{\lambda _{t}\big(\partial \widehat{\sigma }^{t}(\eta ),\partial 
\widehat{\sigma }^{t}(\eta )\big)-2i\,\beta _{t}\big(\partial \widehat{%
\sigma }^{t}(\eta ),\mathbf{s}\eta \big)+\nu (\mathbf{s}\eta ,\mathbf{s}\eta
)\Big\}\,\mathrm{d}t,
\end{eqnarray*}%
because $\mathrm{d}\widehat{w}_{t}\mathrm{d}\widehat{w}_{t}=\mathrm{d}t$.
Here 
\begin{eqnarray*}
\big\langle\xi ,\partial \widehat{\sigma }^{t}(\eta )\big\rangle &=&{\frac{%
\mathrm{d}}{\mathrm{d}\varepsilon }}\,\widehat{\sigma }^{t}(\eta
+\varepsilon \xi )|_{\varepsilon =0}, \\
\lambda _{t}(\eta ,\eta ) &=&\langle 2\func{Re}\zeta _{t},\eta \rangle
^{2},\qquad \nu _{t}(\mathbf{s}\eta ,\mathbf{s}\eta )=\langle \func{Im}\zeta
_{t},\mathbf{s}\eta \rangle ^{2}, \\
\beta _{t}(\eta ,i\xi ) &=&\langle 2\func{Re}\zeta _{t},\eta \rangle
\,\langle \func{Im}\zeta _{t},\xi \rangle \equiv \langle i\boldsymbol{\beta }%
_{t}\xi ,\eta \rangle
\end{eqnarray*}

This gives a quasilinear stochastic equation of the first order for $%
\widehat{\sigma }^{t}(\eta )$: 
\begin{eqnarray*}
\mathrm{d}\widehat{\sigma }^{t} &+&\big\{s(\eta ,\upsilon _{t})+\langle i%
\widetilde{\boldsymbol{\kappa }}_{t}\mathbf{s}\eta ,\partial \widehat{\sigma 
}^{t}\rangle -{\frac{1}{2}}\,\widetilde{\varepsilon }_{t}(\mathbf{s}\eta ,%
\mathbf{s}\eta )\big\}\,\mathrm{d}t \\
&=&\big\{\langle 2\func{Re}\zeta _{t},\partial \widehat{\sigma }^{t}\rangle
+\langle \func{Im}\zeta _{t}^{\sharp },\mathbf{s}\eta \rangle \big\}\,%
\mathrm{d}\widehat{w}_{t}-{\frac{1}{2}}\,\lambda _{t}(\partial \widehat{%
\sigma }^{t},\partial \widehat{\sigma }^{t})\,\mathrm{d}t,
\end{eqnarray*}%
where $\widetilde{\boldsymbol{\kappa }}_{t}=\boldsymbol{\kappa }_{t}-%
\boldsymbol{\beta }_{t}$ and $\widetilde{\varepsilon }_{t}=\varepsilon
_{t}-\nu _{t}$. This equation has a solution in quadratic form 
\begin{equation*}
\sigma ^{t}(\eta )=\ln \widehat{\rho }_{t}+\widehat{\vartheta }_{t}(\eta )+{%
\frac{1}{2}}\,p_{t}(\eta ,\eta ),\qquad \partial \sigma ^{t}(\eta )=\widehat{%
\vartheta }_{t}+\mathbf{p}_{t}\eta ,\qquad \partial ^{2}\sigma ^{t}=\mathbf{p%
}_{t},
\end{equation*}%
where 
\begin{eqnarray*}
\mathrm{d}\ln \widehat{\rho }^{t} &=&\mathrm{d}\widehat{\sigma }%
^{t}(0)=\langle 2\func{Re}\zeta _{t},\widehat{\vartheta }_{t}\rangle \,%
\mathrm{d}\widehat{w}_{t}-{\frac{1}{2}}\,\lambda _{t}(\widehat{\vartheta }%
_{t},\widehat{\vartheta }_{t})\,\mathrm{d}t, \\
\mathrm{d}\widehat{\vartheta }_{t} &=&\mathrm{d}\partial \widehat{\sigma }%
^{t}(0)=(2\mathbf{p}_{t}\func{Re}\zeta _{t}+\mathbf{s}\func{Im}\zeta _{t})\,%
\mathrm{d}\widehat{w}_{t}-\big\{(\mathbf{p}_{t}\boldsymbol{\lambda }_{t}-i%
\mathbf{s}\boldsymbol{\kappa }_{t}^{\intercal })\,\widehat{\vartheta }_{t}-%
\mathbf{s}\upsilon _{t}\big\}\,\mathrm{d}t \\
\mathrm{d}\mathbf{p}_{t} &=&\mathrm{d}\partial ^{2}\widehat{\sigma }^{t}=-%
\big\{\mathbf{p}_{t}\boldsymbol{\lambda }_{t}\mathbf{p}_{t}+\mathbf{s}%
\boldsymbol{\varepsilon }_{t}\mathbf{s}+i\,(\mathbf{p}_{t}\widetilde{%
\boldsymbol{\kappa }}_{t}\mathbf{s}-\mathbf{s}\widetilde{\boldsymbol{\kappa }%
}_{t}^{\intercal }\mathbf{p}_{t})\big\}\,\mathrm{d}t,
\end{eqnarray*}%
and $\widetilde{\kappa }^{\intercal }(\mathbf{p}\eta ,\mathbf{s}\eta )=%
\widetilde{\kappa }(\mathbf{s}\eta ,\mathbf{p}\eta )$. By explicit
expression of the finite-dimensional operators $\boldsymbol{\beta }_{t}$, $%
\boldsymbol{\lambda }_{t}$ and $\boldsymbol{\nu }_{t}$ in terms of $\zeta
_{t},\zeta _{t}^{\sharp }$ , the stochastic integral $\ln \widehat{\rho }%
^{t}=\int_{0}^{t}\mathrm{d}\widehat{\sigma }^{t}(0)$ can be represented in
the form given by theorem~3. The differentials $\mathrm{d}\partial \widehat{%
\sigma }^{t}(0)$ and $\mathrm{d}\partial ^{2}\sigma ^{t}$ can be written in
the form of the stochastic differential equation (\ref{eq:diff4.7}) for $%
\widehat{\vartheta }_{t}(\eta )=\langle \eta ,\widehat{\vartheta }%
_{t}\rangle $ , and in the form of the Riccati equation (\ref{eq:diff4.8})
for $p_{t}(\eta ,\eta )=\langle \eta ,\mathbf{p}_{t}\eta \rangle $ , with $%
\mathbf{k}_{t}=\mathbf{p}_{t}+(i/2)\,\mathbf{s}$ because $\mathbf{s}%
^{\intercal }=-\mathbf{s}$ . Hence 
\begin{eqnarray*}
\langle \eta ,2\mathbf{p}_{t}\func{Re}\zeta _{t} &+&\mathbf{s}\func{Im}\zeta
_{t}\rangle =2\func{Re}\Big\langle\eta ,\Big(\mathbf{p}_{t}+{\frac{i}{2}}\,%
\mathbf{s}\Big)\,\zeta _{t}^{\sharp }\Big\rangle,\qquad \eta \in \func{Re}%
\Xi ^{0}, \\
\lambda _{t}(\mathbf{p}_{t}\eta ,\mathbf{p}_{t}\eta ) &-&2i\beta _{t}(%
\mathbf{p}_{t}\eta ,\mathbf{s}\eta )+\nu _{t}(\mathbf{s}\eta ,\mathbf{s}\eta
)=\big(2\func{Re}\langle \eta ,\mathbf{k}_{t}\zeta _{t}^{\sharp }\rangle %
\big)^{2}.
\end{eqnarray*}%
Theorem 3 is proved.

\noindent \textbf{Remark 3\/} The quantum linear filtering equation (\ref%
{eq:diff4.7}), (\ref{eq:diff4.8}) can be written in the form of the
classical Kalman-Bucy filter 
\begin{eqnarray}
\mathrm{d}\vartheta _{t}(\omega )+(\boldsymbol{\alpha }_{t}^{\intercal
}\vartheta _{t}(\omega )+\mathbf{s}\upsilon _{t})\,\mathrm{d}t &=&2\func{Re}(%
\mathbf{k}_{t}\zeta _{t}^{\sharp })\,\mathrm{d}w_{t},  \notag \\
{\frac{\mathrm{d}}{\mathrm{d}t}}\,\mathbf{p}_{t}+\boldsymbol{\alpha }%
_{t}^{\intercal }\mathbf{p}_{t}+\mathbf{p}_{t}\boldsymbol{\alpha }_{t}+%
\mathbf{s}\widetilde{\boldsymbol{\varepsilon }}_{t}\mathbf{s} &=&\mathbf{p}%
_{t}\boldsymbol{\lambda }_{t}\mathbf{p}_{t},  \label{eq:diff4.9}
\end{eqnarray}%
where $\boldsymbol{\alpha }_{t}=\boldsymbol{\lambda }_{t}\mathbf{p}_{t}+i\,%
\widetilde{\boldsymbol{\kappa }}_{t}\mathbf{s}$, $\boldsymbol{\alpha }%
_{t}^{\intercal }=\mathbf{p}_{t}\boldsymbol{\lambda }_{t}-i\,\mathbf{s}%
\widetilde{\boldsymbol{\kappa }}^{\intercal }$, $\boldsymbol{\lambda }_{t}=4%
\func{Re}\zeta _{t}\func{Re}\zeta _{t}^{\intercal }$. The solution of this
system of equations with the initial $\vartheta _{0}\left( \omega \right)
=\vartheta $, $\mathbf{p}_{0}=\mathbf{1}$ gives $\widehat{\sigma }^{t}(\eta
) $ in the form of the integral 
\begin{eqnarray}
\widehat{\sigma }^{t}(\eta ) &=&\vartheta (\widehat{\eta }_{0})+{\frac{1}{2}}%
\,\widehat{\eta }_{0}^{2}+\int_{0}^{t}\Big\{\langle \mathbf{s}\widehat{\eta }%
_{r},\mathrm{d}\Upsilon _{r}\rangle  \notag \\
&+&{\frac{1}{2}}\,\big[\widetilde{\varepsilon }_{r}(\mathbf{s}\eta _{r},%
\mathbf{s}\eta _{r})-\lambda _{r}\big(\widehat{p}_{r}(\widehat{\eta }_{r}),%
\widehat{p}_{r}(\widehat{\eta }_{r})\big)\big]\,\mathrm{d}r\Big\},
\label{eq:diff4.10}
\end{eqnarray}%
over the stochastic trajectories $\widehat{\eta }_{r}=\widehat{\phi }%
_{r}(t,\eta )$ of the adjoint equation (\ref{eq:diff4.4}) with $\widehat{p}%
_{r}=\widehat{\vartheta }_{r}+\mathbf{p}_{r}\widehat{\eta }_{r}$, $\widehat{%
\eta }_{0}=\widehat{\phi }_{0}(t,\eta )$, and 
\begin{equation*}
\mathrm{d}\Upsilon _{t}=\upsilon _{t}\mathrm{d}t+\func{Im}\zeta _{t}^{\sharp
}\mathrm{d}\widehat{w}_{t}.
\end{equation*}%
Indeed, if $\mathrm{d}_{-}\widehat{\eta }_{r}=\widehat{\eta }_{r}-\widehat{%
\eta }_{r-\mathrm{d}r}$ is the backward stochastic differential, then $%
\mathrm{d}\langle \widehat{\vartheta }_{r},\widehat{\eta }_{r}\rangle
=\langle \widehat{\eta }_{r},\mathrm{d}\widehat{\vartheta }_{r}\rangle
+\langle \widehat{\vartheta }_{r},\mathrm{d}_{-}\widehat{\eta }_{r}\rangle $%
, and 
\begin{equation*}
\mathrm{d}\,\big[p_{r}(\widehat{\eta }_{r},\widehat{\eta }_{r})\big]=2p_{r}(%
\widehat{\eta }_{r},\mathrm{d}_{-}\widehat{\eta }_{r})+\dot{p}_{r}(\widehat{%
\eta }_{r},\widehat{\eta }_{r})\,\mathrm{d}r.
\end{equation*}%
Using the equations (\ref{eq:diff4.9}) and writing the equation (\ref%
{eq:diff4.4}) in the form $\mathrm{d}_{-}\widehat{\eta }_{r}=\boldsymbol{%
\alpha }_{r}\widehat{\eta }_{r}\mathrm{d}r-2\func{Re}\zeta _{r}\mathrm{d}%
\widehat{u}_{r}$ with respect to 
\begin{equation*}
\mathrm{d}\widehat{u}_{r}=\mathrm{d}\widehat{w}_{r}+\langle \mathbf{p}_{r}%
\widehat{\eta }_{r},2\func{Re}\zeta _{r}\rangle \,\mathrm{d}r,
\end{equation*}%
one can obtain, by integrating by parts, the difference $\widehat{\sigma }%
^{t}(\eta )-\ln \widehat{\rho }^{t}-\vartheta (\eta _{0})-{\frac{1}{2}}\,%
\widehat{\eta }_{0}^{2}$: 
\begin{eqnarray*}
\widehat{\vartheta }_{t}(\eta ) &-&\vartheta (\widehat{\eta }_{0})+{\frac{1}{%
2}}\,p_{t}(\eta ,\eta )-{\frac{1}{2}}\,\widehat{\eta }_{0}^{2} \\
&=&\int_{0}^{t}\Big\{\langle \widehat{\eta },\mathrm{d}\widehat{\vartheta }%
\rangle +{\frac{1}{2}}\,\dot{\mathbf{p}}\widehat{\eta }\mathrm{d}r-\langle
d_{-}\widehat{\eta },\widehat{p}(\widehat{\eta })\rangle \Big\} \\
&=&\int_{0}^{t}\Big\{\langle \mathbf{s}\widehat{\eta },\mathrm{d}\Upsilon
\rangle -\langle 2\func{Re}\zeta \mathrm{d}\widehat{w},\widehat{\vartheta }%
\rangle +\langle \boldsymbol{\alpha }\widehat{\eta },\mathbf{p}\widehat{\eta 
}\rangle +{\frac{1}{2}}\,\dot{(}\widehat{\eta },\widehat{\eta })-\lambda (%
\mathbf{p}\widehat{\eta },\mathbf{p}\widehat{\eta })\,\mathrm{d}r\Big\} \\
&=&\int_{0}^{t}\Big\{\langle \mathbf{s}\widehat{\eta },\mathrm{d}\Upsilon
\rangle -\hat{\vartheta}(2\func{Re}\zeta )\,\mathrm{d}\widehat{w}+{\frac{1}{2%
}}\,\big[\widetilde{\varepsilon }(\mathbf{s}\widehat{\eta },\mathbf{s}%
\widehat{\eta })+\lambda (\widehat{\vartheta },\widehat{\vartheta })-\lambda
(\widehat{p},\widehat{p})\big]\Big\} \\
&=&\int_{0}^{t}\Big\{\langle \mathbf{s}\widehat{\eta },\mathrm{d}\Upsilon
\rangle +{\frac{1}{2}}\,\big[\varepsilon (\mathbf{s}\widehat{\eta },\mathbf{s%
}\widehat{\eta })-\lambda \big(\widehat{p},\widehat{p}\big)\big]\,\mathrm{d}r%
\Big\}-\ln \widehat{\rho }^{t},
\end{eqnarray*}%
which gives (\ref{eq:diff4.10}) with $\ln \widehat{\rho }^{t}=\int_{0}^{t}(%
\widehat{\vartheta }(2\func{Re}\zeta )\,\mathrm{d}\widehat{w}-{\frac{1}{2}}%
\,\lambda (\widehat{\vartheta },\widehat{\vartheta })\,\mathrm{d}r)$.

\noindent \textbf{Example.\/} Let us consider the quantum linear filtering
equations (\ref{eq:diff4.9}) for the case of indirect nondemolition
observation of the coordinate of free quantum Brownian motion, described in
the above example. Since $\zeta _{t}=\zeta _{t}^{\sharp }$ we obtain $%
\boldsymbol{\beta }_{t}=0$, $\boldsymbol{\gamma }_{t}=0$, $\boldsymbol{\nu }%
_{t}=0$, and $\boldsymbol{\lambda }_{t}=\lambda \left( 
\begin{array}{cc}
0 & 0 \\ 
0 & 1%
\end{array}%
\right) =4\boldsymbol{\varepsilon }_{t}$. This allows us to define the
unique positive 
\begin{equation*}
\mathbf{p}_{t}={\frac{1}{\func{Re}\omega _{t}}}\,\left( 
\begin{array}{cc}
2\,|\omega _{t}|^{2} & \func{Im}\omega _{t} \\ 
\func{Im}\omega _{t} & 1/2\,%
\end{array}%
\right) ={\frac{1}{2}}\,(\mathbf{k}_{t}+\mathbf{k}_{t}^{\sharp }),
\end{equation*}
solution of the Riccati equation (\ref{eq:diff4.9}) with the initial $%
\mathbf{p}_{0}=1$. Here%
\begin{equation*}
\mathbf{k}_{t}={\frac{1}{\func{Re}\omega _{t}}}\,\left( 
\begin{array}{cc}
2\,\,|\omega _{t}|^{2} & i\omega _{t}^{\ast } \\ 
-i\omega _{t} & 1/2\,%
\end{array}%
\right) ,\;\;\;\mathbf{k}_{t}^{\sharp }={\frac{1}{\func{Re}\omega _{t}}}%
\,\left( 
\begin{array}{cc}
2\,\,|\omega _{t}|^{2} & -i\omega _{t} \\ 
i\omega _{t}^{\ast } & 1/2\,%
\end{array}%
\right) ,
\end{equation*}%
correspond to the degenerate form $k_{t}(\eta ^{\sharp },\eta )=|2\omega
_{t}\eta _{p}+i\eta _{q}|^{2}/(2\func{Re}\omega _{t})$ of the solution 
\begin{equation*}
{\frac{\mathrm{d}}{\mathrm{d}t}}\,\mathbf{k}_{t}+\mathbf{k}_{t}\boldsymbol{%
\lambda }_{t}\mathbf{k}_{t}=i\bigg\{\mathbf{s}\,\bigg({\frac{1}{2}}\,%
\boldsymbol{\lambda }_{t}+i\boldsymbol{\omega }_{t}\bigg)\,\mathbf{k}_{t}+%
\mathbf{k}_{t}\bigg({\frac{1}{2}}\,\boldsymbol{\lambda }_{t}-i\boldsymbol{%
\omega }_{t}\bigg)\,\mathbf{s}\bigg\}
\end{equation*}%
with the initial $\mathbf{k}_{0}=\mathbf{1}+(i/2)\mathbf{s}$ and fixed $%
\mathbf{k}_{t}-\mathbf{k}_{t}^{\sharp }=i\mathbf{s}$. In addition, the
parameter $\omega _{t}\in \mathbf{C}$, that defines the a posteriori wave
function 
\begin{equation*}
\widehat{\psi }^{t}(x)=\big(\sqrt{\func{Re}\omega _{t}/\pi }\,\widehat{\rho }%
^{t}\big)^{1/2}\exp \bigg\{-{\frac{1}{2}}\,\omega _{t}(x-\widehat{q}%
_{t})^{2}+{\frac{i}{2}}\,\widehat{p}_{t}x\bigg\}
\end{equation*}%
for $\widehat{\varphi }^{t}=(\widehat{\psi }^{t}|X(\xi )\,\widehat{\psi }%
^{t})$ satisfies the one-dimensional complex Riccati equation 
\begin{equation*}
{\frac{\mathrm{d}}{\mathrm{d}t}}\,\omega _{t}+{\frac{i}{2m}}\,\omega
_{t}^{2}={\frac{1}{2}}\,\lambda ,\qquad \omega _{0}={\frac{1}{2}}\,,
\end{equation*}%
which has (only) one positive solution 
\begin{equation*}
\omega _{t}={\frac{\alpha }{2}}\,{\frac{1+\alpha \mathrm{th}(\lambda
t/\alpha )}{\mathrm{th}(\lambda t/\alpha )+\alpha }},\qquad \alpha =\sqrt{%
\lambda m/2}\,(1-i).
\end{equation*}%
The a posteriori mathematical expectations 
\begin{eqnarray*}
\widehat{p}_{t} &=&{\frac{1}{\widehat{\rho }^{t}}}\,\int \widehat{\psi }%
^{t}(x)^{\ast }{\frac{2}{i}}\,{\frac{\mathrm{d}}{\mathrm{d}x}}\,\widehat{%
\psi }^{t}(x)\,\mathrm{d}x={\frac{1}{\widehat{\rho }^{t}}}\,(\widehat{\psi }%
^{t}\mid P\widehat{\psi }^{t}), \\
\widehat{q}_{t} &=&{\frac{1}{\widehat{\rho }^{t}}}\,\int \widehat{\psi }%
^{t}(x)^{\ast }x\,\widehat{\psi }^{t}(x)\,\mathrm{d}x={\frac{1}{\widehat{%
\rho }^{t}}}\,(\widehat{\psi }^{t}\mid Q\widehat{\psi }^{t}),
\end{eqnarray*}%
of the momentum and the coordinate of the quantum Brownian particle satisfy
the linear system of stochastic equations 
\begin{eqnarray*}
\mathrm{d}\widehat{p}_{t}+\sqrt{\lambda }\,\displaystyle{\frac{\func{Im}%
\omega _{t}}{\func{Re}\omega _{t}}}\,\mathrm{d}\widehat{w}_{t}=0,\hfill
&&\quad \widehat{p}_{0}=p, \\
\mathrm{d}\widehat{q}_{t}-\displaystyle{\frac{1}{m}}\widehat{p}_{t}\mathrm{d}%
t=\displaystyle{\frac{i}{2}}\sqrt{\lambda }\,\displaystyle{\frac{1}{\func{Re}%
\omega _{t}}}\,\mathrm{d}\widehat{w}_{t},\hfill &&\quad \widehat{q}_{0}=q.
\end{eqnarray*}%
The solution of this system gives the probability density 
\begin{equation*}
\widehat{\rho }^{t}=\exp \bigg\{\int_{0}^{t}\bigg(\sqrt{\lambda }\,\widehat{q%
}_{r}\,\mathrm{d}\widehat{w}_{r}-{\frac{1}{2}}\,\widehat{q}_{r}^{2}\mathrm{d}%
r\bigg)\bigg\}
\end{equation*}%
for the indirect nondemolition observation $Y(t)$ of the coordinate of the
free quantum Brownian particle with the mass $m$. Note, that the solution $%
\omega _{t}$ tends exponentially to the limit $\omega _{\infty }=\alpha $
which defines the finite a posteriori dispersions 
\begin{eqnarray*}
\sigma _{t}^{2} &=&{\frac{\big\|(P-\widehat{p}\,I)\,\widehat{\psi }_{0}^{t}%
\big\|^{2}}{\Vert \widehat{\psi }_{0}^{t}\Vert ^{2}}}=2\,{\frac{|\omega
_{t}|^{2}}{\func{Re}\omega _{t}}} \\
\tau _{t}^{2} &=&{\frac{\big\|(Q-\widehat{q}\,I)\,\widehat{\psi }_{0}^{t}%
\big\|^{2}}{\Vert \widehat{\psi }_{0}^{t}\Vert ^{2}}}={\frac{1}{2\func{Re}%
\omega _{t}}}
\end{eqnarray*}%
and the correlation $\rho _{t}=\Vert \widehat{\psi }_{0}^{t}\Vert -2\func{Re}%
((P-\widehat{p}\,I)\,\widehat{\psi }_{0}^{t}|(Q-\widehat{q}\,I)\,\widehat{%
\psi }_{0}^{t})$ in the limit $t\rightarrow \infty $: 
\begin{equation*}
\sigma _{\infty }^{2}=\sqrt{2\lambda m},\qquad \tau _{\infty }^{2}=\sqrt{%
2/(\lambda m)},\qquad \rho _{\infty }=-1.
\end{equation*}

It is well known that the unobserved free quantum particle becomes `fuzzy'
and has an infinite a posteriori limit dispersions. The contradiction of the
localised motion of the quantum particle, which is seen during the
continuous observation, and of the impossibility of localising this particle
on the basis of the von Neumann or Lindblad equation, has provided a source
for quantum paradoxes such as the Zeno paradox \cite{bib:diff4}. The derived
quantum filtering (\ref{eq:diff2.6}) and a posteriori von Neumann (\ref%
{eq:diff2.10}) equations resolves these quantum paradoxes not only on the
qualitative level, but also on the quantitative level of the microscopic
quantum stochastic model for the continuous observation.

\section{Appendix}

\renewcommand{\theequation}{{\rm A}.\arabic{equation}}

\setcounter{equation}{0}

\textbf{1.} \ Let $\{\mathrm{H}_\xi |\xi >1\}$ be a continuous family of
Hilbert subspaces $\mathrm{H}_\xi \subseteq \mathrm{H}$ with nondecreasing
norms: $\eta \leq \xi \Leftrightarrow \Vert \psi \Vert _\eta \leq \Vert \psi
\Vert _\xi $, $\forall \,\psi \in \mathrm{H}$. It will be called a \textit{%
scale of the Hilbert space} $\mathrm{H}$ with a scalar product $(\psi |\psi
)=\lim _{\xi \downarrow 1}\Vert \psi \Vert _\xi ^2$. An \textit{inductive
limit} $\lim _{\xi \downarrow 1}\mathrm{H}_\xi $ of the Hilbert scale $\{%
\mathrm{H}_\xi \}$ is defined as a pre-Hilbert space $\mathrm{D}=\cup _{\xi
>1}\mathrm{H}_\xi $, provided with the inductive convergence: 
\begin{equation*}
\psi _n\to 0\Leftrightarrow \exists \,\xi >1\,:\,\Vert \psi _n\Vert _\xi \to
0.
\end{equation*}
The operator $X:\mathrm{D}\to \mathrm{D}$ is called (\textit{inductively}) 
\textit{continuous} if $X\psi _n\to 0$ for any convergent sequence $\{\psi
_n\}$ that tends to zero. This means, that the restriction of $X$ to any
subspace $\mathrm{H}_\zeta \in \{\mathrm{H}_\xi \}$ is a continuous map into
some $\mathrm{H}_\xi \subseteq \mathrm{D}$, i.e. for any $\zeta >1$ there
exists $\xi >1$, such that 
\begin{equation}  \label{eq:diffA.1}
\Vert X\Vert _\zeta ^\xi :=\sup _{\psi \in \mathcal{H}_\zeta }\bigg\{{\frac{%
\Vert X\psi \Vert _\xi }{\Vert \psi \Vert _\zeta }}\bigg\}<\infty .
\end{equation}
The set $\mathcal{B}(\mathrm{D})$ of all continuous operators $X$ , having
the Hermitian-conjugate operators $X^{*}$ on the pre-Hilbert space $\mathrm{D%
}$ form an associative algebra with the identity $I\in \mathcal{B}(\mathrm{D}%
)$ and the involution $X^{**}=X$. This algebra is a $C^{*}$-algebra of the
bounded operators only if $\mathrm{D}=\mathrm{H}$.

\noindent \textbf{2.} \ Let $\Gamma _{+}$ be the set of all the chains $\tau
_{n}=(t_{1},\ldots ,t_{n})$, $t_{i}\in \mathbf{R}_{+}$, $t_{1}<\cdots <t_{n}$
of length $n<\infty $, identified with finite subsets $\tau \subset \mathbf{R%
}_{+}$, $\tau =\{t_{1},\ldots ,t_{n}\}$ of the cardinality $|\tau |=n\in
\{0,1,\ldots \}$. We denote by $\mathrm{d}\tau =\prod_{t\in \tau }\mathrm{d}%
t $ the positive $\sigma $-finite measure on $\Gamma _{+}=\sum_{n=0}^{\infty
}\Gamma _{n}$, which is defined as a sum $\sum_{n=0}^{\infty }\mathrm{d}\tau
_{n}$ of the measures $\mathrm{d}\tau _{n}=\mathrm{d}t_{1}\cdots \mathrm{d}%
t_{n}$ on $\Gamma _{n}=\{|\tau |=n\}$ with the only atom $\mathrm{d}\tau
_{0}=1$ on the empty chain $\tau _{0}=\emptyset $ corresponding to $|\tau
|=0 $. The Hilbert space $\mathrm{H}=L^{2}(\Gamma _{+})$ of the
square-integrable functions $f:\Gamma _{+}\rightarrow \mathbf{C}$, $%
(f|f)<\infty $, where $(f|f)=\int |f(\tau )|^{2}\mathrm{d}\tau $, 
\begin{equation}
\int g(\tau )\,\mathrm{d}\tau :=\sum_{n=0}^{\infty
}\didotsint\limits_{0<t_{1}<\cdots <t_{n}<\infty }g(t_{1},\ldots ,t_{n})\,%
\mathrm{d}t_{1}\cdots \mathrm{d}t_{n},  \label{eq:diffA.2}
\end{equation}%
is naturally identified with the \textit{Fock space} $\mathcal{F}=\Gamma (%
\mathcal{K}) $ , of the sequences $f=\{\varphi _{n}|n=0,1,\ldots \}$ of
symmetric continuations $\varphi _{n}:\mathbf{R}_{+}^{n}\rightarrow \mathbf{C%
}$ of the functions $f(\tau _{n})$ , with the scalar product 
\begin{equation*}
(f\mid f)=\sum_{n=0}^{\infty }{\frac{1}{n!}}\int_{0}^{\infty }\cdots
\int_{0}^{\infty }\big|\varphi (t_{1},\ldots ,t_{n})\big|^{2}\mathrm{d}%
t_{1}\cdots \mathrm{d}t_{n}.
\end{equation*}%
The Hilbert scale $\{\mathcal{F}_{\xi }|\xi >1\}$ of the dense subspaces 
\begin{equation*}
\mathcal{F}_{\xi }=\bigg\{f\in \mathcal{F}\,\mid \,\Vert f\Vert _{\xi
}^{2}=\int \xi ^{|\tau |}\big|f(\tau )\big|^{2}\mathrm{d}\tau <\infty \bigg\}
\end{equation*}%
is called Fock \cite{bib:diff9}--\cite{bib:diff11} over the Hilbert space $%
\mathcal{K}=L^{2}(\mathbf{R}_{+})$. The function $f(\tau )=\delta
_{\emptyset }(\tau )$ ( where $\delta _{\emptyset }(\tau )=1$ for $\tau
=\emptyset $, $\delta _{\emptyset }(\tau )=0$ for $\tau =\emptyset $),
normalised $(\Vert f\Vert _{\xi }^{2}=\int \xi ^{|\tau |}\delta _{\emptyset
}(\tau )\,\mathrm{d}\tau =1)$ with respect to any $\xi $, is called the 
\textit{vacuum} function.

\noindent \textbf{3.} \ The most important examples of the continuous
operators, and their conjugates $\mathcal{F}_{+}\to \mathcal{F}_{+}$ on the
inductive limit $\mathcal{F}_{+}=\cup _{\xi >1}\mathcal{F}_\xi $, are the
`quantum annihilation' operators $\widehat{a}_t$ on the interval $[0,t)$,
which act as the integrals 
\begin{equation*}
[\widehat{a}_tf]\,(\tau )=\int_0^tf(\tau \sqcup r)\,\mathrm{d}r,\qquad
\forall \,f\in \mathcal{F}.
\end{equation*}
Here $\tau \sqcup r$ is a chain, defined almost everywhere $(r\not \in \tau
) $ as a union $\{\tau ,r\}$ of the single-point chain $r\in \mathbf{R}_{+}$
with some chain $\tau =\{t_1,\ldots ,t_n\}$. Although the operators $%
\widehat{a}_t $ possess a complete system of exponential eigenfunctions 
\begin{equation*}
f(\tau )=k^{\otimes }(\tau ):=\prod_{t\in \tau }k(t),\qquad k\in l^2(\mathbf{%
R}_{+}),
\end{equation*}
they are not normal, i.e. they do not commute with their adjoint operators $%
\widehat{a}_t^{*}$, defined by the finite sums 
\begin{equation*}
[\widehat{a}_t^{*}f](\tau )=\sum_{r\in \tau }^{r<t}f(\tau \backslash
r),\qquad \forall \,f\in \mathcal{F}_{+},
\end{equation*}
where $\tau \backslash r=\{t\in \tau |t\neq r\}$. Polynomials of the
operators $\{\widehat{a}_t^{*},\widehat{a}_t|t\in \mathbf{R}_{+}\}$ form the 
\textit{Weyl algebra} over the simple functions $g\in L^2(\mathbf{R}_{+})$,
generating the whole algebra $\mathcal{B}(\mathcal{F}_{+})$ through the
triviality of the commutant: 
\begin{equation*}
\big\{Y\in \mathcal{B}(\mathcal{F}_{+})\mid X\in \{\widehat{a}_t^{*},%
\widehat{a}_t\}\Leftrightarrow [X,Y]=0{\big\}}=\mathbf{C}\widehat{1}.
\end{equation*}
Linear elements of the algebra $\mathcal{B}(\mathcal{F}_{+})$, defined as
the quantum Wiener integrals 
\begin{equation*}
\widehat{r}(f,g)=\int_0^\infty \big(f(t)\,\mathrm{d}\widehat{a}_t+g(t)\,%
\mathrm{d}\widehat{a}_t^{*}\big)=\widehat{a}(f)+\widehat{a}^{*}(g)
\end{equation*}
of the square-integrable functions $f,g:\mathbf{R}_{+}\to \mathbf{C}$,
together with the identity operator ${\hat 1}$ , form the *-representation $%
\widehat{r}(f,g)^{*}=\widehat{r}(g^{*},f^{*})$ of the canonical commutation
relations 
\begin{equation}  \label{eq:diffA.3}
\big[\widehat{r}(f,g),\widehat{r}(f,g)^{*}\big]=\big(\Vert f\Vert ^2-\Vert
g\Vert ^2\big)\,\hat 1
\end{equation}
with respect to the involution $(f,g)^{\sharp }=(g^{*},f^{*})$. The vacuum
function $\delta _\emptyset \in \mathcal{F}$ induces a Gaussian state on $%
\mathcal{B}(\mathcal{F}_{+})$, defined by the characteristic functional 
\begin{equation*}
\big(\delta _\emptyset \mid e^{i\widehat{r}(f,g)}\delta _\emptyset \big)%
=\exp \Big\{{\frac 12}\,\big(\Vert f\Vert ^2-\Vert g\Vert ^2\big)\Big\}
\end{equation*}
on the Hilbert space of pairs $\eta =(f,g)$ with the norm$^2$ $|\eta |^2={\
\frac 12}\,(\Vert f\Vert ^2+\Vert g\Vert ^2)$ and symplectic bilinear form $%
s(\eta ^{\prime },\eta )$ which equals $s(\eta ^{\sharp },\eta )={\frac 1i}%
\,(\Vert f\Vert ^2-\Vert g\Vert ^2)$ for $\eta ^{\prime }=(g^{*},f^{*})$.

\noindent \textbf{4.} \ The measurable function $F:t\mapsto F(t)$ whose
values are the operators $\mathcal{D}\to \mathcal{D}$ of the inductive limit 
$\mathcal{D}=\cup _{\xi >1}\mathcal{H}_\xi $ is called locally $p$-\textit{%
integrable in the inductive scale} $\{\mathcal{H}_\xi =\mathrm{H}_\xi
\otimes \mathcal{F}_\xi \}$ , if, for any $\zeta >1,$ there exists a $\xi >1$%
, such that 
\begin{equation*}
\Vert F\Vert _{\zeta ,t}^{\xi ,p}:=\bigg(\int_0^t\big(\Vert F(r)\Vert _\zeta
^\xi \big)^p\mathrm{d}r\bigg)^{1/p}<\infty ,\qquad \forall \,t\in \mathbf{R}%
_{+}.
\end{equation*}
In particular, this condition means that the operators $F(t):\mathcal{D}\to 
\mathcal{D}$ are continuous for almost all $t\in \mathbf{R}_{+}$. For such
square-integrable functions the quantum stochastic integrals $\widehat{a}%
_t^{*}(F)$, $\widehat{a}_t(F)$ on $\{\widehat{a}_r^{*}|r<t\}$ and $\{%
\widehat{a}_r|r<t\}$ are defined to be the operators $\mathcal{D}\to 
\mathcal{D}$: 
\begin{eqnarray}
\big[\widehat a_t^*(F)\,h\big](\tau)&=&\sum_{r\in \tau}^{r<t}\big[F(r)\,h%
\big]\, (\tau),\qquad\forall\, h\in\mathcal{D},  \notag \\
\big[\widehat a_t(F)\,h\big](\tau)&=&\int_0^t\big[F(r)\,\dot h(r)\big]\,
(\tau).  \label{eq:diffA.4}
\end{eqnarray}
In this equation $h\mapsto \dot h(r)$ is a Maliven derivative, defined in
the Fock representation $h:\Gamma _{+}\rightarrow \mathrm{D}$ of the
elements $h\in \mathcal{D}$ almost everywhere $(r\not \in \tau )$ by the
vector-function $\dot h(r)\in \mathcal{D\ }$as $\dot h(r,\tau )=h(r\sqcup
\tau )$. The continuity of the operators $\widehat{a}_t(F)$, $\widehat{a}%
_t^{*}(F)$ in $\mathcal{D}$ follows directly from the estimates 
\begin{eqnarray*}
\big\|\widehat a_t(F)\big\|_{\zeta+\varepsilon}^\xi&\leq&\sqrt{1/\varepsilon}%
\,\|F\|_{\zeta,t}^{\xi,2}, \qquad\forall\,\varepsilon>0, \\
\big\|\widehat a_t^*(F)\big\|_{\zeta}^{\xi-\varepsilon}&\leq&\sqrt{%
\xi/\varepsilon}\, \|F\|_{\zeta,t}^{\xi,2}, \qquad\forall\,\varepsilon<\xi.
\end{eqnarray*}
obtained in \cite{bib:diff5}, \cite{bib:diff9}--\cite{bib:diff11}. From
these estimates, if there exists $\xi >1$ for any $\zeta >1$ such that $%
\Vert F\Vert _{\zeta ,t}^{\xi ,2}$, $\Vert D\Vert _{\zeta ,t}^{\xi
,2}<\infty $ then, for any $\zeta ^{+}>1$ , there exists a $\xi _{-}>1$, for
which the operator 
\begin{equation*}
\imath _0^t(F,D)=\widehat{a}_t(F)+\widehat{a}_t^{*}(D)
\end{equation*}
is bounded from $\mathcal{H}_{\zeta ^{+}}$ to $\mathcal{H}_{\xi _{-}}$. To
be precise, by choosing for every $\zeta ^{+}>1$ a $\zeta >1$, such that $%
\zeta <\zeta ^{+}$, and an $\varepsilon >\xi -1$, such that $\varepsilon
<\zeta ^{+}-\zeta $, we obtain 
\begin{equation*}
\big\|\imath _0^t(F,D)\big\|_{\zeta ^{+}}^{\xi _{-}}\leq \sqrt{\xi
/\varepsilon }\,\big( \Vert F\Vert _{\zeta ,t}^{\xi ,2}+\Vert D\Vert _{\zeta
,t}^{\xi ,2}\big)
\end{equation*}
for any $\xi _{-}$ on the nonempty interval $(1,\xi -\varepsilon ]$.

\noindent \textbf{5.} \ If $D(t)\in \mathcal{B}(\mathcal{D})$ for almost all 
$t\in \mathbf{R}_{+}$ and the adjoint function $D^{*}(t)=D(t)^{*}$ is also
locally integrable on $\mathcal{D}$, then $\widehat{a}_t^{*}(D)\in \mathcal{B%
}(\mathcal{D})$ and $\widehat{a}_t^{*}(D)^{*}=\widehat{a}_t(D^{*})$. This
means that the integral $\imath _0^t(D^{*},D)$ is (formally) self-adjoint.
Moreover, the set of integrals 
\begin{equation*}
X(t)=\imath _0^t(D^{*},D)+\int_0^tG(r)\,\mathrm{d}r,
\end{equation*}
where $G:\mathbf{R}_{+}\to \mathcal{B}(\mathcal{D})$ is locally integrable $%
(p=1)$, together with the $G^{*}$ , function, forms a $*$-algebra with
respect to the pointwise operator product $(X^{*}X)(t)=X(t)^{*}X(t)$. This
product is defined by the quantum nonadapted Ito formula \cite{bib:diff10} 
\begin{eqnarray}
X(t)^*X(t)&=&\imath_0^t(D^*X+X^*D,F^*D+X^*D)  \notag \\
&+&\int_0^t[G^*X+X^*_rD+D^*D+D^*X_r+X^*G](r)\,\mathrm{d}r,\qquad
\label{eq:diffA.5}
\end{eqnarray}
which corresponds to the case of the locally square-integrable operator
function $t\mapsto X_t(t)$, $X_t^{*}(t)=X(t)_t^{*}$. Here $X\mapsto X_t$
means a derivation, which is defined by%
\begin{equation*}
[X_th](\tau )=[Xh]\,(\tau \sqcup t)-\big[X\dot h(t)\big](\tau )
\end{equation*}
for almost all $\tau \in \Gamma $, $t\notin \tau $. In this instance, the
case $X_t(t)=0=X_t^{*}(t)$ for all $t\in \mathbf{R}_{+}$ corresponds to the
adaptive property of the operator functions $F(t),D(t)$ and $G(t)$ .

Comprehensive information on explicit stochastic integration in the Fock
scale is given in \cite{bib:diff20}.\bigskip\ 

\noindent \textbf{Acknowledgements}\medskip\ 

The author expresses his gratitude to A. S. Holevo and A.~N.~Shiriaev for
useful discussions of the article and helpful remarks.

\end{document}